\documentclass[fleqn,usenatbib]{mnras}

\usepackage[T1]{fontenc}

\DeclareRobustCommand{\VAN}[3]{#2}
\let\VANthebibliography\thebibliography
\def\thebibliography{\DeclareRobustCommand{\VAN}[3]{##3}\VANthebibliography}

\usepackage{graphicx}	% Including figure files
\usepackage{amsmath}	% Advanced maths commands
\usepackage{amssymb}	% Extra maths symbols
\usepackage{soulutf8}
\usepackage{bm}
\usepackage{comment}
\usepackage{multirow}

\usepackage{newtxtext,newtxmath}

\title[Oscillating magnetised hybrid stars]{Oscillating magnetised hybrid stars under the magnifying glass of multi-messenger observations}
\author[M. Mariani et al.]
{Mauro Mariani$^{1,2}$\thanks{E-mail: mmariani@fcaglp.unlp.edu.ar},
Lucas Tonetto$^{3,4}$,
M. Camila Rodríguez$^{1,2}$,
Marcos O. Celi$^{1}$,
\newauthor
Ignacio F. Ranea-Sandoval$^{1,2}$,
Milva G. Orsaria$^{1,2}$
and Aurora Pérez Martínez$^{5}$\thanks{On  leave from Instituto de Cibernética, Matemática y Física ICIMAF, Calle E esq a 15 No 309, Vedado La Habana 10400, Cuba.}
\\
$^{1}$Grupo de Gravitación, Astrofísica y Cosmología, Facultad de Ciencias Astronómicas y Geofísicas, Universidad Nacional de La Plata,\\Paseo del Bosque S/N, 1900, Argentina\\
$^{2}$CONICET, Godoy Cruz 2290, 1425 Buenos Aires, Argentina\\
$^{3}$Dipartimento di Fisica, ``Sapienza'' University of Rome, Piazzale A. Moro, 5. 00185 Roma, Italy\\
$^{4}$INFN, Sezione di Roma, Piazzale A. Moro, 5. 00185 Roma, Italy\\
$^{5}$Departamento de Física Fundamental, Universidad de Salamanca, Plaza de la Merced s/n 37008, Spain
}

\date{Accepted XXX. Received YYY; in original form ZZZ}

\pubyear{2015}

\begin{document}
\label{firstpage}
\pagerange{\pageref{firstpage}--\pageref{lastpage}}
\maketitle

\begin{abstract}
We model neutron stars as magnetised hybrid stars with an abrupt hadron-quark phase transition in their cores, taking into account current constraints from nuclear experiments and multi-messenger observations. We include magnetic field effects considering the Landau level quantisation of charged particles and the anomalous magnetic moment of neutral particles. We construct the magnetised hybrid equation of state, and we compute the particle population, the matter magnetisation and the transverse and parallel pressure components. We integrate the stable stellar models, considering the dynamical stability for \emph{rapid} or \emph{slow} hadron-quark phase conversion. Finally, we calculate the frequencies and damping times of the fundamental and $g$ non-radial oscillation modes. The latter, a key mode to learn about phase transitions in compact objects, is only obtained for stars with slow conversions. For low magnetic fields, we find that one of the objects of the GW170817 binary system might be a hybrid star belonging to the slow extended stability branch. For magnetars, we find that a stronger magnetic field always softens the hadronic equation of state. Besides, only for some parameter combinations a stronger magnetic field implies a higher hybrid star maximum mass. Contrary to previous results, the incorporation of anomalous magnetic moment does not affect the studied astrophysical quantities. We discuss possible imprints of the microphysics of the equation of state that could be tested observationally in the future, and that might help infer the nature of dense matter and hybrid stars.
\end{abstract}


%%%%%%%%%%%%%%%%% BODY OF PAPER %%%%%%%%%%%%%%%%%%

\section{Introduction}

Neutron stars (NSs) are found to have very strong magnetic fields (MFs), $10^8$-$10^{15}$~Gauss at their surface; the extreme $10^{13}$-$10^{15}$~Gauss interval is associated to a particular kind of NSs, the so-called \emph{magnetars}. Up to date, thirty two magnetars and candidates are known, thirty enumerated in the McGill Online Magnetar Catalog \citep{Olausen:2014tmm} and two recently discovered \citep{Coti:2021tnm, Huang:2021sgo}. Magnetars are characterised by high electromagnetic activity, mostly at energy scales of X-ray and soft $\gamma$-ray, such as soft gamma repeaters (SGRs) and anomalous X-ray pulsars (AXPs). The phenomenology associated to the surface of these objects may be explained due to the evolution and decay of the MF and its disrupting effects on the NS crust \citep{Kaspi:2017mar}. In the interior of NSs, however, the strength, distribution and effects of the MF are still unknown. This uncertainty is added to the fact that the structure and internal composition of NSs are also open questions. While we understand better the behaviour of matter at densities below the nuclear saturation density, $n_0 \approx 0.16$~fm$^{-3}$, the description of matter in NS cores, whose densities can traditionally reach up to $\sim 10 \ n_0$, is not well determined.

The matter in NSs is in the high-density, low-temperature regime of the QCD phase diagram, where the theory predicts the occurrence of a phase transition from hadrons to deconfined quarks. This turns NSs naturally impressive laboratories to study and determine the \emph{equation of state} (EoS) of dense matter. There exist various proposals about the composition of these objects (see, for example, \cite{APR1998, Norsen:2002sn, Baldo:2003ns, Peng:2008dp, Zdunik:2013mm, Baym:2018fht, McLerran:2019qm, Orsaria:2019pti, Tanimoto:2020mn}). One of the most discussed hypotheses is the presence of quark matter in their innermost regions, giving rise to \emph{hybrid stars} (HSs). To confirm the presence of quark matter in NSs, the combination of different available astrophysical data should allow us to distinguish quantitative changes in the material properties of their dense matter composition \citep{Alford:2019sfq}. Recently, the theoretical (model-independent) evidence of quark-matter cores in NSs has been suggested by \citet{Annala:2020efq}.

On the other hand, the detection of gravitational waves (GWs) emitted by NS mergers, along with its electromagnetic counterpart, has marked a milestone in multi-messenger astronomy. A known mechanism capable of emitting GWs \citep[see, for example,][and references therein]{Glampedakis:2018GWF} that might be detectable in the next generation of GW detectors are non-radial pulsations of isolated NSs \citep{anderssonferrari2011}. One of the most promising scenarios in which such non-radial perturbations might be detected are the hot and perturbed leftovers of binary NS mergers, the proto-NSs. Other astronomical events in which proto-NSs copiously emit GWs are Type-II Supernovae \citep[see, for example,][and references therein]{sagert:2009PhRvL,Radice:2019ApJL}. These hot and perturbed proto-NSs radiate energy through neutrinos and GWs to achieve their equilibrium state \citep[see][and references therein]{Tsang:2019mtp,most:2019PhRvL,camelio2017}.  Therefore, the analysis of non-radial oscillation modes can provide us information about the internal composition of these compact objects \citep{Lasky:2015gwf,camelio2017}. Such oscillations are called quasi-normal modes (QNMs), and have complex frequencies whose real and imaginary parts represent the pulsation frequency and damping time, respectively. The damping time gives a measurement of how fast the amplitude of a given oscillation reduces its original amplitude by a factor $e$, and is associated to the energy stored in a given mode and the rate of energy loss due to emission of GWs.

Fluid QNMs of NSs can be classified according to the corresponding restoring force: the high-frequency $p$-modes -for which the pressure is the main restoring force-, the low-frequency $g$-modes -in which buoyancy causes fluid elements to oscillate-, and the fundamental $f$-mode -whose frequency values are between the previous two families and it is expected to be the dominant emitter-. Particularly, in this work we analyse NSs that have already reached their equilibrium state but still emit GWs. Our focus will be on the $g$-mode associated with first-order hadron-quark phase transition. A detection of a $g$-mode emitted from a cold non-rotating NS (for which other $g$-modes are not present) could be used to determine properties of the hadron-quark phase transition \citep{minuitti2003,sotani2001PhRvD,Ranea:2018omo}.

In addition to the modes mentioned above, it is known that extremely magnetised compact objects present a different family of modes in their crusts, known as \emph{torsional} modes 
\cite[see, for example,][]{sotani2007MNRAS,Asai2014}. This kind of modes can be excited, for example, as a consequence of a giant flare \cite[see][and references therein]{levin2011MNRAS}. In this work, we are interested in exploring the interior of these compact objects, so we will skip the study of \emph{torsional} modes. Furthermore, the fundamental and first overtone torsional modes are expected to have frequencies below the kHz, and should be distinguishable from the ones we are interested in.

Microscopically, the presence of the MF produces the quantisation of the transverse moment of the electrically charged particles into \emph{Landau levels}. Besides the Landau quantisation, the MF also interacts with particles through their \emph{anomalous magnetic moment} (AMM). The AMM couples to the MF and produces a Zeeman effect in the energy spectrum of the particles, suppressing the spin degeneracy \citep{Tsai:1971moc}. The term that involves the AMM in the energy spectrum of fermions can be approximated as a linear contribution in the field if the MF is weak ($B << m^2$, where $m$ is the naked mass of fermions), thus for neutral particles this is a valid approximation. For strong MF and charged fermions, however, the  existence of a critical MF, accounting for the validity of such approximation, must be taken into account for a consistent treatment of the magnetised EoS including AMM \citep{Ferrer:2015iot}. While for low MF the AMM does not play any role, for high MF ($B > 10^{15}$~Gauss), AMM may affect the EoS and should be considered \citep{Broderick:2002eos}. In this work, we will include the AMM of non-charged particles, following  the paper of \citet{Ferrer:2015iot}, where it has been demonstrated that only the contribution of the AMM from neutral particles has to be included in the EoS, regardless of the strength of the MF. 

From the observational point of view, the very accurate mass detections of the pulsars PSR J1614-2230 \citep{Demorest:2010ats} and PSR J0348+0432 \citep{Antoniadis:2013amp} (with $M \sim 2$~M$_\odot$) have started a new era for the NS research field. Since then, there have been an ever increasing number of measurements and detections related to NSs. The mass measurement of PSR J1614-2230 was improved by \citet{Arzoumanian:2018tny}, and, recently, a new high mass pulsar was detected, PSR J0740+6620 \citep{Cromartie:2020rsd}, also later corrected by \citet{Fonseca:2021rfa}. In addition, the GW observatories have confirmed two NS mergers, GW170817 \citep{Abbott:2017oog} and GW190425 \citep{Abbott:2020goo}, and two black hole-NS mergers, GW200105 and GW200115. In particular, GW170817 also had the electromagnetic counterpart detected, GRB170817A and AT2017gfo \citep{Abbot:2017gwa, Abbott:2017mmo}. These observations constraint not only the mass and radius of NSs, but also their dimensionless  tidal deformability, $\Lambda$; assuming that both merging objects have the same EoS and spin period values according to the galaxy observed ones, the dimensionless tidal deformability of a $1.4$~M$_\odot$ object was estimated to be in the range of $70$-$580$, at $90 \%$ confidence level \citep{Abbott:2018exr}. Last but not least, the start up of NICER detector has provided particularly relevant observations in the last years; the collaboration has measured the mass and radius of the isolated pulsars PSR J0030+0451 \citep{Miller:2019pjm,Riley2019anv} and, in a joint observation with XMM-Newton, PSR J0740+6620 \citep{Miller:2021tro,Riley:2021anv}. The measured value of the radius of this last observation, obtained with a $\sim 15 \%$ error at a confidence level of $68\%$, introduces a slight tension with previous observations: for PSR J0740+6620 ($M \sim 2~M_{\odot}$) it was obtained a radius very similar to the estimated radius of PSR J0030+0451 ($M \sim 1.4~M_{\odot}$), even though they have very different masses.

This work is a continuation of a previous paper, \citet{Mariani:2019mhs}, but with the improvement of incorporating AMM effects, the analysis of the tidal deformability on the $\Lambda_1$-$\Lambda_2$ plane and the non-radial oscillations calculations, focusing in the fundamental mode -which is expected to be the mode in which more energy is channelled into GWs- and the $g$-mode -related to the existence of a sharp hadron-quark phase transition in the inner core of such compact objects-. Furthermore, we use an improved magnetised crust in the stellar hybrid configurations and we developed a new hadronic parametrisation taking into account the recent astrophysical constraints by NICER. For convenience, we use the natural units system, $c=\hbar=1$. Details and definitions of electromagnetic and MF natural units are presented in the Appendix of our previous paper.

This paper is organised as follows. In Section~\ref{eos}, we present the microphysical considerations and the corresponding results; we describe the model used, considering the presence of the MF and its effects over the thermodynamics of the system and we present the thermodynamic expressions for the different phases of the hybrid hadron-quark EoS. We also present the construction method used for a sharp phase transition and discuss the treatment of the MF anisotropy. In Section~\ref{structure}, we present the stellar structure equations and the formalism employed to analyse the dynamic stability of stellar configurations when slow and rapid phase conversions are considered at the quark-hadron interface. We also show the results obtained for magnetised hybrid stars. In Section~\ref{oscillations}, we present the non-radial oscillation modes theory and the results obtained for the frequencies and damping times of the $f$ and $g$-modes. A summary of the work, discussion about the astrophysical implications and conclusions are provided in Section~\ref{conclus}. 

\section{Magnetised Hybrid EoS}
\label{eos}

\subsection{Magnetic Field treatment}

The strong MF in NSs evolves and dissipates not only in their crust but also in the inner layers and core of these objects. Using numerical magneto-hydrodynamic simulations, it has been found that a stable MF configuration is obtained from the combination of a toroidal and a poloidal MF \citep{braithwaite2006evolution,Ciolfi:2013ttc,Sur:2020mfc}. This chaotic MF distribution complicates the choice of a suitable and realistic profile of MF if one thinks of a simple scheme to solve the HSs structure equations. In general, it is usual to use a functional form to model the MF profile in a given direction. \cite{Dexheimer2017} use a polynomial MF profile in the star's polar direction, satisfying Maxwell's equations. However, it is not clear the validity of this phenomenological adjustment in a direction other than polar. A universal polynomial MF profile in magnetars has also been suggested by \cite{Oertel2019}. This universal functional form implies very high MF ($\sim 5 \times 10^{17}$~Gauss) on the surface of the star and an almost flat profile of the MF, which can generate instabilities in the matter composing NSs \citep{Thapa:2020eos}.

We will use a monotonically increasing towards the centre of the star exponential profile to parametrise the MF strength inside the star. It is adjusted to produce MF values at the star surface compatible with those observed in magnetars (see McGill Online Magnetar Catalog\footnote{http://www.physics.mcgill.ca/~pulsar/magnetar/main.html}). Moreover, in order to prevent the HSs phase transition from inducing a discontinuity in the effective MF, the adopted profile will be a function of the baryonic chemical potential, in the following form:
\begin{equation} \label{param}
	B(\mu_b) = B_{\text{min}}+B_{\text{max}} \left[ 1-\mathrm{e}^{\beta\frac{(\mu_b-m_n)^\alpha}{m_n}} \right] \,
\end{equation}
where $\alpha=2.5$ and $\beta = -4.08 \times 10^{-4}$ and $m_n$ is the nucleon mass \citep{Dexheimer:2012hsi}. The $\alpha$ and $\beta$ are adjusted to reproduce the values of a MF parametrised in terms of the density from \citet{Dexheimer:2012hsi}. The parameters $B_{\text{min}}$ and $B_{\text{max}}$ will be fixed to study two astrophysics scenarios: the regular \emph{low MF} pulsar and the \emph{magnetar}, see Table~\ref{table:bcases}.

Although hypothetical, as the MF behaviour inside NSs is not completely determined, this method is a suitable approximation for ultra strong MFs at magnetar cores, and is widely used in the literature \citep{Bandyopadhyay:1997qmf,Mao:2003aso,Rabhi:2009qhp,Dexheimer:2012hsi,Flores:2020gws,Thapa:2020eos}.

\begin{table} 
	\centering
	\begin{tabular}{cccc} 
		\hline
		Scenario & $B_{\text{min}}$ {[Gauss]} & $B_{\text{max}}$ [Gauss] \\
		\hline
		Low MF & $1 \times 10^{13}$ & $1 \times 10^{15}$ \\
		Magnetar & $1 \times 10^{15}$ & $3 \times 10^{18}$ \\
		\hline
	\end{tabular}
		\caption{MF parametrization values for the two selected astrophysical scenarios.}
		\label{table:bcases}
\end{table}

To consider the effects of the MF on the particles of the system, we take into account the electric charge, $q$, and the AMM, $\kappa$. We will include AMM only for neutral particles, following the work of \citet{Ferrer:2015iot}, where it has been shown that, for electrically charged particles, the AMM has no significant effect on the EoS for any MF strength value. Therefore, without the inclusion of the AMM, in the presence of a $z-$direction MF, the momentum of electrically charged particles is quantized into Landau levels in the perpendicular direction to the MF. The inclusion of Landau levels in the transverse momentum and in the energy spectrum of charged particles is performed in the same way as in \citet{Mariani:2019mhs}.

For electrically neutral particles with AMM $\kappa$, in the presence of a $z-$direction MF, the energy spectrum is
\begin{equation}
	E = \sqrt{k_z^2 + \bar{m}^2}\, ,
\end{equation}
where, in this case,
\begin{equation}
	\bar{m}^2 = \left(\sqrt{m^2 + k_\perp^2} - s \kappa B \right)^2 \, \bm{,}
\end{equation}
being $k_z$ and $k_\perp$ the momentum z-component and transverse component respectively, and $s$ the spin projection of the particle.

As it can be seen, Landau level quantisation does not exist and the interaction with the MF is only through the AMM. Hence, although the AMM participates in the thermodynamic expressions and the sum over spin projections has to be considered, there is no modification of the thermodynamic integrals into sums for any momentum component. In the next subsections, we will detail the specific thermodynamics expressions for each phase of the hybrid EoS. 

\subsection{Hybrid EoS phases}

\subsubsection{Subnuclear regime}
    
To model the magnetised matter at sub-nuclear densities in the crust of the HS, we adopt an EoS from the work of \citet{Mutafchieva:2019rol}. This EoS treats, in a unified and consistent way, both the outer and inner regions of the crust in the framework of the nuclear-energy density functional. As \citet{Mutafchieva:2019rol} detailed, the outer crust is modeled as a layered structured of successive perfect crystals made of a single nuclear species embedded in an electron relativistic Fermi gas. The onset of the inner crust is defined by the drip out of the neutrons from the nuclei. Thus, in the inner crust atomic nuclei coexist with unbound neutrons, also in an electron background. This magnetised crust EoS reaches densities up to $n_B = 0.07$~fm$^{-3}$.

\subsubsection{Hadron phase}

To describe the hadron phase for the outer core of HSs, we use the SW4L parametrisation of the \emph{relativistic mean field} (RMF) model \citep{Spinella:2018dab, Malfatti:2020onm}, where the exchange of $\sigma$, $\omega$, $\rho$, $\sigma^*$ and $\phi$ mesons describe the interaction among the baryons and that includes density dependent meson-baryon coupling constants. Considering a uniform and external magnetic field along the $z$ axis, the Lagrangian is given by
\begin{eqnarray}
  \mathcal{L} &=& \sum\limits_b \overline\psi_b\bigl[\gamma_{\mu}
    (i\partial^{\mu}-g_{\omega b}\omega^{\mu}-g_{\phi
      b}\phi^{\mu}-\tfrac{1}{2}g_{\rho b}(n)\boldsymbol{\tau}\cdot
    \boldsymbol{\rho}^{\mu})\nonumber\\ &&-(m_b-g_{\sigma
      b}\sigma-g_{\sigma^*
      b}\sigma^*-\frac{1}{2} \kappa_b \sigma_{\mu\nu} F^{\mu\nu})\bigr]\psi_b\nonumber\\
  &&+\tfrac{1}{2}\left(\partial_{\mu}\sigma\partial^{\mu}\sigma
  -m^2_{\sigma}\sigma^2\right)\nonumber\\ &&-\tfrac{1}{3}b_{\sigma}m_n\left(g_{\sigma
    N}\sigma\right)^3 -\tfrac{1}{4}c_{\sigma}\left(g_{\sigma
    N}\sigma\right)^4\nonumber\\ &&-\tfrac{1}{4}\omega_{\mu\nu}\omega^{\mu\nu}
  +\tfrac{1}{2}m^2_{\omega}\omega_{\mu}\omega^{\mu}\nonumber\\ &&
  -\tfrac{1}{4}\boldsymbol{\rho}_{\mu\nu}\cdot\boldsymbol{\rho}^{\mu\nu}\nonumber
  +\tfrac{1}{2}m^2_{\rho}\boldsymbol{\rho}_{\mu}\cdot\boldsymbol{\rho}^{\mu}\\
  &&-\tfrac{1}{4}\phi^{\mu\nu}\phi_{\mu\nu}+\tfrac{1}{2}m^2_{\phi}\phi_{\mu}
  \phi^{\mu}\nonumber\\ &&+\tfrac{1}{2}\left(\partial_{\mu}
  \sigma^*\partial^{\mu}\sigma^* -m^2_{\sigma^*}\sigma^{*2}\right)-\frac{1}{4} F_{\mu\nu}F^{\mu\nu}\, ,
 \label{laghad}
\end{eqnarray}
where the sum over $b$ takes account of the baryon octet and $\Delta$ resonances. Scalar ($\sigma, ~\sigma^*$), vector ($\omega,~ \phi$), and isovector ($\rho$) meson fields mediate the interactions among baryons. $F_{\mu\nu}=\partial_{\mu}A_{\nu}-\partial_{\nu}A_{\mu}$ is the electromagnetic field tensor and $\sigma_{\mu\nu}=\frac{i}{2} [\gamma_{\mu}, \gamma_{\nu}]$. The strength of the AMM is $\kappa_b=0$ for charged baryons and  $\kappa_b \neq 0$ for neutral baryons. In particular, the AMM of the $\Delta^0$ neutral baryon is not yet fully determined (see, for example, the works by \citet{Cloet:2003dbm, Aubin:2009lco}; we will use the value provided in the recent work by \citet{Machavariani:2011cca}. In Table~\ref{table:amm}, we present the AMM values for all the neutral baryons. Note that AMM arises from effective theory for neutral fermions and it can be thought of as an additional coupling constant. For charged fermions, the AMM appears as a radiative correction to the theory and it does not significantly affect the EoS \citep{Ferrer:2015iot}.  

In the SW4L parametrization, only the meson-baryon couplings of the $\rho$-meson, $g_{\rho b}(n)$, depend on the local baryon number density in the following way
\begin{equation}
g_{\rho b}(n) = g_{\rho b}(n_0)\,\mathrm{exp}\left[\,-a_{\rho}
  \left(\frac{n_B}{n_0} - 1\right)\,\right] \, ,
\end{equation}
where $n_B = \sum_b n_b$ is the total baryon number density. The other meson-baryon couplings are constants adjusted consistently.
\begin{table}
\begin{center}
\begin{tabular}{|c|c|}
\hline 
Neutral baryon & AMM, $\kappa_b / \mu_N$ \\ \hline
$n$ & -1.91 \\
$\Lambda$ & -0.61 \\
$\Sigma^0$ & 1.61 \\
$\Xi^0$ & -1.25 \\
$\Delta^0$ & -2.50 \\
\hline
\end{tabular}
\caption{AMM for the considered electrically neutral baryons. The AMM is given in terms of the nuclear magneton, $\mu_N=q_e/2m_N$. The values are from Particle Data Group \citep{Zyla:2020zbs}, except for $\Delta^0$ that is from the work by \protect\citet{Machavariani:2011cca}.}
\label{table:amm}
\end{center}
\end{table}

By using the RMF approximation to solve the Euler-Lagrange equations of motion that follow from Eq.~(\ref{laghad}), we obtain the system of coupled equations given by
\begin{eqnarray}
m_{\sigma}^2 \bar{\sigma} &=& \sum_{b} g_{\sigma b} n_b^s -
\tilde{b}_{\sigma} \, m_N\,g_{\sigma N} (g_{\sigma N} \bar{\sigma})^2
\nonumber\\ & & - \tilde{c}_{\sigma} \, g_{\sigma N} \, (g_{\sigma N}
\bar{\sigma})^3 \, \nonumber\\ m_{\sigma^*}^2 \bar{\sigma^*} &=&
\sum_{b} g_{\sigma^* b} n_b^s\, , \nonumber\\ m_{\omega}^2 \bar{\omega}
&=& \sum_{b} g_{\omega b} n_{b}\, , \\ m_{\rho}^2\bar{\rho} &=&
\sum_{B}g_{\rho B}(n)I_{3b} n_{b} \, , \nonumber\\ m_{\phi}^2
\bar{\phi} &=& \sum_{b} g_{\phi b} n_{b}\, , \nonumber
\end{eqnarray}
where $I_{3b}$ is the third component of isospin, and the scalar $n_b^s$ and particle number density $n_{b}$ for neutral baryons are given by \citep{Broderick:2000teo}
\begin{eqnarray}
n_{b}^s &=&  \frac{m_b^*}{4 \pi^2} \sum_s k_{\rm{F}}^{b} E_{\rm{F}} - \bar{m}_b^2
\ln \left( \left| \frac{ E_{\rm{F}} + k_{\rm{F}}^{b}} {\bar{m}_b}\right|
\right) \, ,
\label{ns_neut}
\end{eqnarray}
\begin{eqnarray}
n_b &=& \frac{1}{2 \pi^2}  \sum_s \frac{1}{2} s \kappa_b B \left[ \bar{m}_b k_{\rm{F}}^{b} + E_{\rm{F}}^{b 2} \left(
\arcsin \frac{\bar{m}_b}{E_{\rm{F}}^b} - \frac{\pi}{2} \right) \right]\nonumber\\
&+& \frac{1}{3} k_{\rm{F}}^{b 3}\,,
\label{n_neut}
\end{eqnarray}
where the total Fermi momentum, $k_{\rm{F}}^{b} = \sqrt{E_{\rm{F}}^b - m_b^{*2}-\sigma \kappa_b B}$. For charged baryons, the scalar and particle number density are the same as in \citet{Mariani:2019mhs}. The spin degeneracy factor is $\gamma_s=2$ for spin $1/2$ particles and $\gamma_s=4$ for spin $3/2$ particles. The effective masses and the Fermi energy are given by
\begin{align*}
m_b^*&= m_b - g_{\sigma b}\bar{\sigma}-g_{\sigma^* b}\bar{\sigma^*} \, ,\nonumber \\ 
\bar{m}^2_{b}& = \left( \sqrt{m_b^{*2} + 2 \nu | q_b | B}- s \kappa_b B \right)^2 \, ,\nonumber\\
E_{\rm{F}}^b &= \mu_b-g_{\omega b} \bar{\omega} - g_{\rho b}(n) \bar{\rho} I_{3B} - g_{\phi b} \bar{\phi} - \widetilde{R} \, , \nonumber
\end{align*}
where $\widetilde{R} = [\partial g_{\rho b}(n)/\partial n] I_{3b} n_b \bar{\rho}$ (see the work by \citet{Malfatti:2020onm} for details). In this work, we use the parameter set and coupling constants given in the work by \citet{Malfatti:2020onm} for SW4L. For meson--$\Delta$ coupling constants we use a quasi-universal choice 
\begin{equation}
x_{{\sigma} {\Delta}} = x_{{\omega} {\Delta}} = 1.1 \,, ~~  x_{{\rho}
  {\Delta}} =x_{{\phi} {\Delta}} =1.0\,, ~~ x_{{\sigma^*} {\Delta}} =0.0\, ,
\end{equation}
where we have considered the value of meson--$\Delta$ coupling relative to that of the nucleon for j = $\sigma, \omega, \rho, \phi$, being $x_{j \Delta} = g_{j\Delta}/g_{j N}$, and $x_{\sigma^* \Delta} = g_{\sigma^*\Delta}/g_{\sigma^* \Lambda}$ with $g_{\sigma^* \Lambda}$ = 1.9242.

The energy density for neutral baryons can be expressed as
\begin{eqnarray}
\epsilon_b &=& \frac{1}{4 \pi^2} \sum_s \frac{1}{2} E_{\rm{F}}^{b 3} k_{\rm{F}}^{b}
 + \frac{2}{3} s \kappa_b B E_{\rm{F}}^{b 3} \left(
\arcsin \frac{\bar{m}_b}{E_{\rm{F}}^{b}} - \frac{\pi}{2} \right) \nonumber \\
&+& \left( \frac{1}{3} s \kappa_b B - \frac{1}{4} \bar{m}_b \right) \left[
\bar{m}_b k_{\rm{F}}^{b} E_{\rm{F}}^{b} + \bar{m}_b^3 \ln \left(\left| \frac{ E_{\rm{F}}^{b} + k_{\rm{F}}^{b}}
{\bar{m}_b}\right| \right) \right]\,,
\label{eb_n}
\end{eqnarray}
the energy density for charged baryons is the same as in \citet{Mariani:2019mhs}, and the total energy density is given by,
\begin{align}
\epsilon_B =& \sum_b \epsilon_{b} + \frac{1}{2} \left( m_{\sigma}^2\bar{\sigma}^2 + m_{\omega}^2 \bar{\omega}^2 + m_{\rho}^2 \bar{\rho}^2 + m_{\sigma^*}^2
\bar{\sigma^*}^2 +  m_{\phi}^2
\bar{\phi}^2 \right) \nonumber \\
 &+ \frac{1}{3} b_{\sigma} m_N g_{\sigma N} \bar{\sigma}^3 + \frac{1}{4} c_{\sigma} m_N g_{\sigma N} \bar{\sigma}^4 \, .
 \label{enden}
 \end{align}
 Using Eq.~(\ref{enden}), the anisotropic pressures can be expressed as
 \begin{equation}
P_{\parallel} = \sum_b \mu_b n_b - \epsilon_B \, ,
\label{ppareq}
\end{equation}
\begin{equation}
P_{\perp} = \sum_b \mu_b n_b - \epsilon_B - \mathcal{M} B \, ,
\label{ppereq}
\end{equation}
where $\mathcal{M}$ is the magnetisation, which we obtained numerically from Eq.~(\ref{magmatter}), and can be neglected, as we will show later.

\begin{figure}
    \centering
    \includegraphics[width=0.95\linewidth]{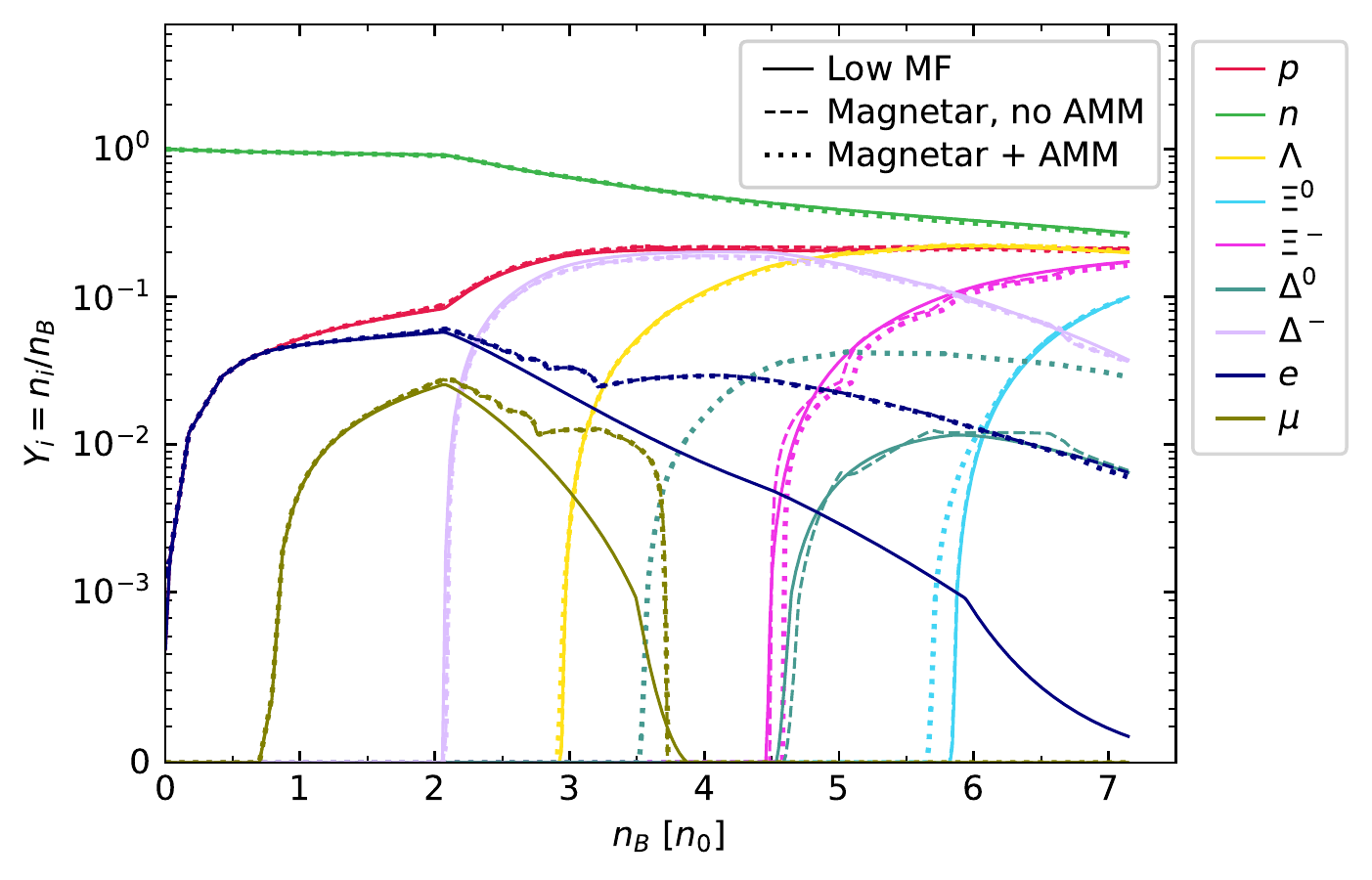}
     \caption{Hadronic particle population, $Y_i=n_i/n_B$, as a function of the baryon number density, $n_B$, in units of the saturation density $n_0$, for different MF scenarios. Strong MF has a predominant effect on the lepton fraction -electrons and muons- which increases as the magnetic field becomes more intense. Particle population of $\Delta^0$ is strongly affected by AMM, lowering its onset density. The inclusion of AMM has no significant influence on the rest of the particle populations.}
    \label{fig:popH}
\end{figure}

\subsubsection{Quark phase}

To describe the quark phase in the inner core of the HSs, we assume deconfined quarks $u$, $d$, $s$ within the Field Correlator Method (FCM) model \citep{Simonov:2007dtf, Simonov:2007vpt}.  In the zero temperature regime, the FCM could be treated as a two free parameter model, the gluon condensate, $G_2$, and the large distance static $\bar{q}q$ potential, $V_1$. As we will show later, we explore the $G_2$-$V_1$ parameter space to select representative sets of parameters that conform to current multi-messenger constraints. We construct the quark EoS considering the Landau levels quantisation and no AMM, because quarks have non-zero electric charge. Thus, the magnetised quark EoS is obtained following the same procedure as in \citet{Mariani:2019mhs}.

\subsubsection{Leptons}

In the case of cold deleptonised HSs, the only leptons present are electrons and muons in both hadron and quark phases. We treat leptons as a free Fermi magnetised gas without the incorporation of AMM, since electrons and muons are electrically charged particles. Their contribution to the magnetised lepton EoS is obtained considering the same expressions as in the case of quark matter, setting the degeneracy factor $\gamma_c=1$ and the parameters $V_1 = G_2 = 0$.

\begin{figure}
    \centering
    \includegraphics[width=0.85\linewidth]{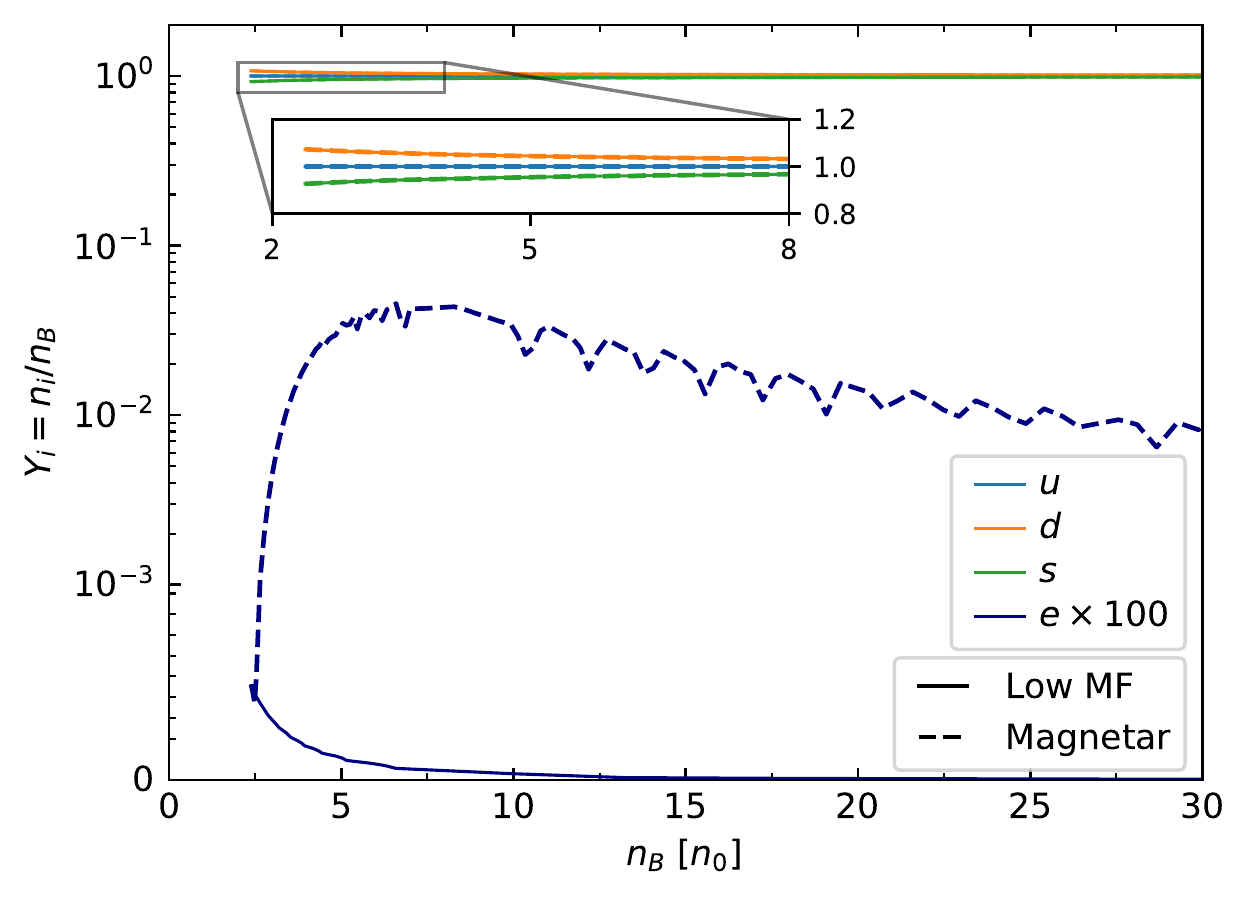}
    \caption{Quark particle population, $Y_i=n_i/n_B$, as a function of the baryon number density, $n_B$, in units of the saturation density $n_0$, for the different MF scenarios and a representative FCM set of parameters. In the enlarged box, the $u$, $d$, $s$ quark fractions do not present noticeable changes with the variation of the MF. The electron population increases significantly for the magnetar case. The oscillatory behaviour is associated to the Landau quantization. Muons are not present in any case.
    }
    \label{fig:popQ}
\end{figure}

\subsection{Magnetised Hybrid EoS construction}

In both phases, we consider baryon number conservation, electric charge neutrality and $\beta$-equilibrium condition.

Local baryon number conservation is given by
\begin{equation}
    \sum_i q_{b,i} n_i = n_B \, ,
\end{equation}
and local charge neutrality by
\begin{equation}
    \sum_i q_{e,i} n_i + \sum_l q_{e,l} n_l = 0 \, ,
\end{equation}
where the sums over $i$ takes account of the species of particles in each phase, baryons or quarks, and the sum over $l$ consider the leptons, $n_i$ and $n_l$ are the respective particle number densities, $q_{b,i}$ is the baryon charge and $q_{e,i},q_{e,l}$ the electric charge of each particle, and $n_B$ is the total baryon number density.

The $\beta$-equilibrium condition -equilibrium under weak interactions-, establishes for the $i$-particle chemical potential, $\mu_i$,
\begin{equation}
\mu_i = q_{b,i} \, \mu_B - q_{e,i} \, \mu_e \, ,
\end{equation}
where $\mu_B$ is the baryon chemical potential and $\mu_e$ is the electron chemical potential. These chemical equilibrium constraints reduce the numbers of freedom of the system, and allow writing the thermodynamic quantities only in terms of $\mu_B$ and $\mu_e$.

\begin{figure*}
    \centering
    \includegraphics[width=0.45\linewidth,angle=0]{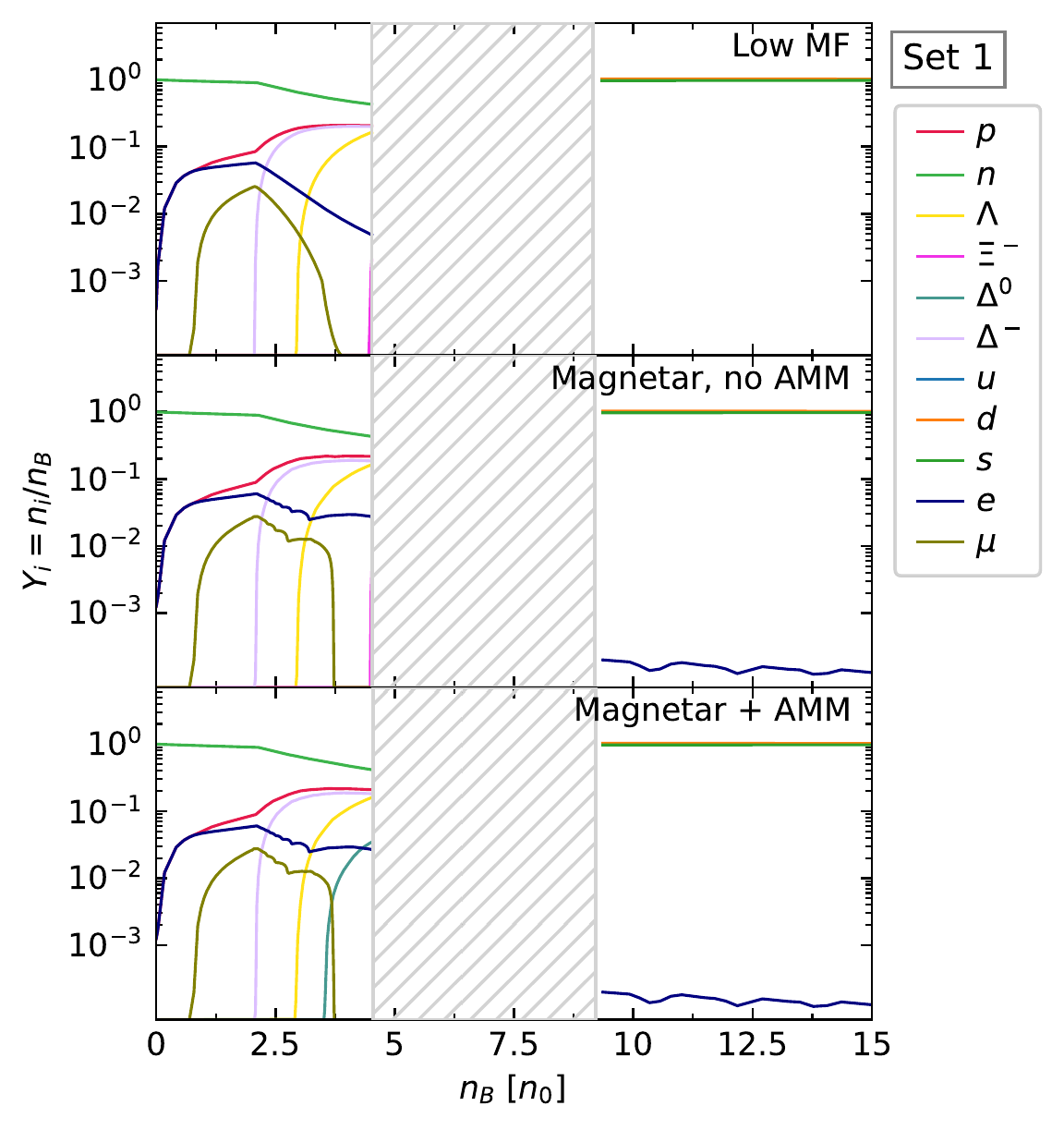}
    \includegraphics[width=0.45\linewidth,angle=0]{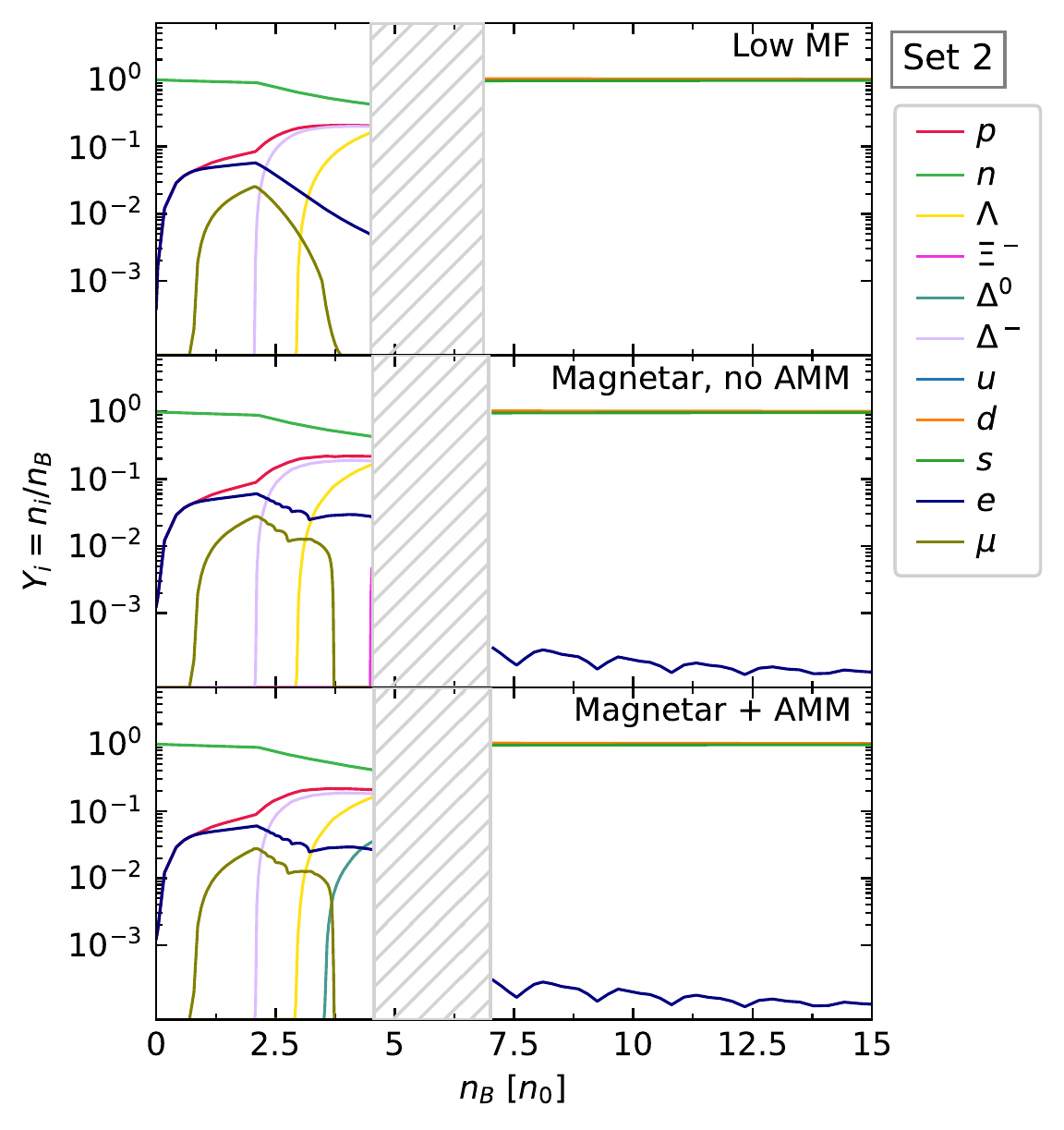}
    \includegraphics[width=0.45\linewidth,angle=0]{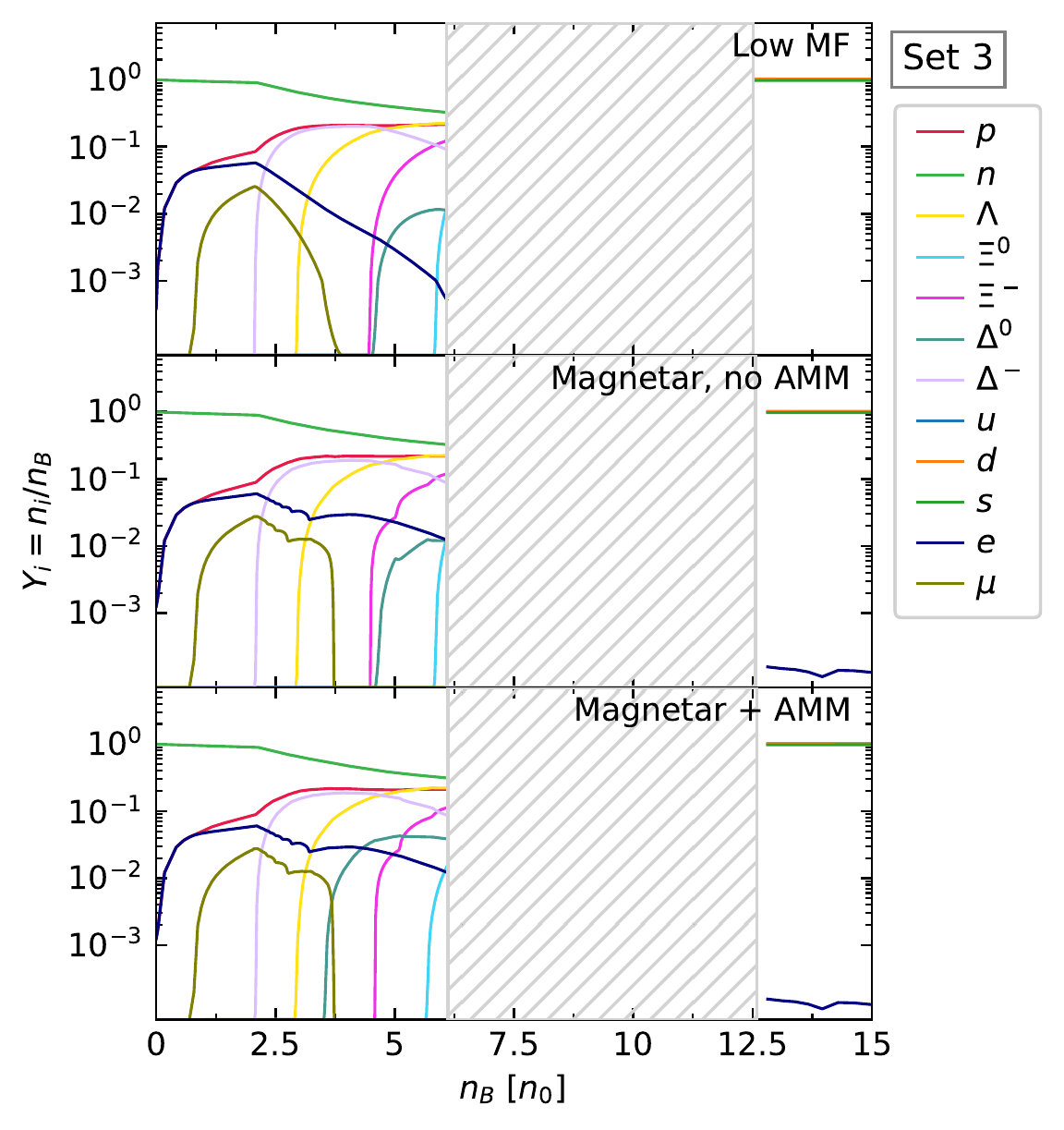}
    \caption{Particle population of the hybrid EoS, $Y_i=n_i/n_B$, as a function of the baryon number density, $n_B$, in units of the saturation density $n_0$, for different MF scenarios and the sets of Table~\ref{table:sets}. The hatched grey area indicates the density jump due to the abrupt phase transition; lower (higher) densities correspond to the hadron (quark) phase. The influence of the MF on the onset and width of the phase transition is not significant, but increasing MF intensity produces noticeable changes in the behaviour of the particle population.}
    \label{fig:popHyb}
\end{figure*}

There are two quantities that define the nature of the hadron-quark phase transition, the surface tension at the hadron-quark interface, $\sigma _{\rm{HQ}}$, and the nucleation timescale of the hadron-quark transition. Both of them are poorly known and have strong astrophysical implications. The surface tension value determines if the phase transition is sharp or smooth  \citep[for a review on the wide spectrum of theoretical values for the hadron-quark surface tension, see][and references therein]{lugonesuniverse7120493}. For large enough values, $\sigma_{\rm{HQ}} > \sigma_{\textrm{crit}}$ ($\sigma_{\textrm{crit}} \sim 70$ MeV fm$^{-2}$) a sharp abrupt transition, with a density discontinuity and no mixed phase region is predicted. This scenario is known as \emph{Maxwell construction}. Otherwise, a smooth \emph{Gibbs} phase transition occurs that includes the formation of geometrical structures in the mixed hadron-quark phase \citep{Voskresensky:2003csa, Endo:2011roh, Wu:2019nse}. Since $\sigma_{\rm{HQ}}$ is a model-dependent quantity, and the smooth or sharp nature of the phase transition is not determined, we adopt the Maxwell formalism as a working hypothesis.

\begin{figure*}
    \centering
    \includegraphics[width=0.45\linewidth,angle=0]{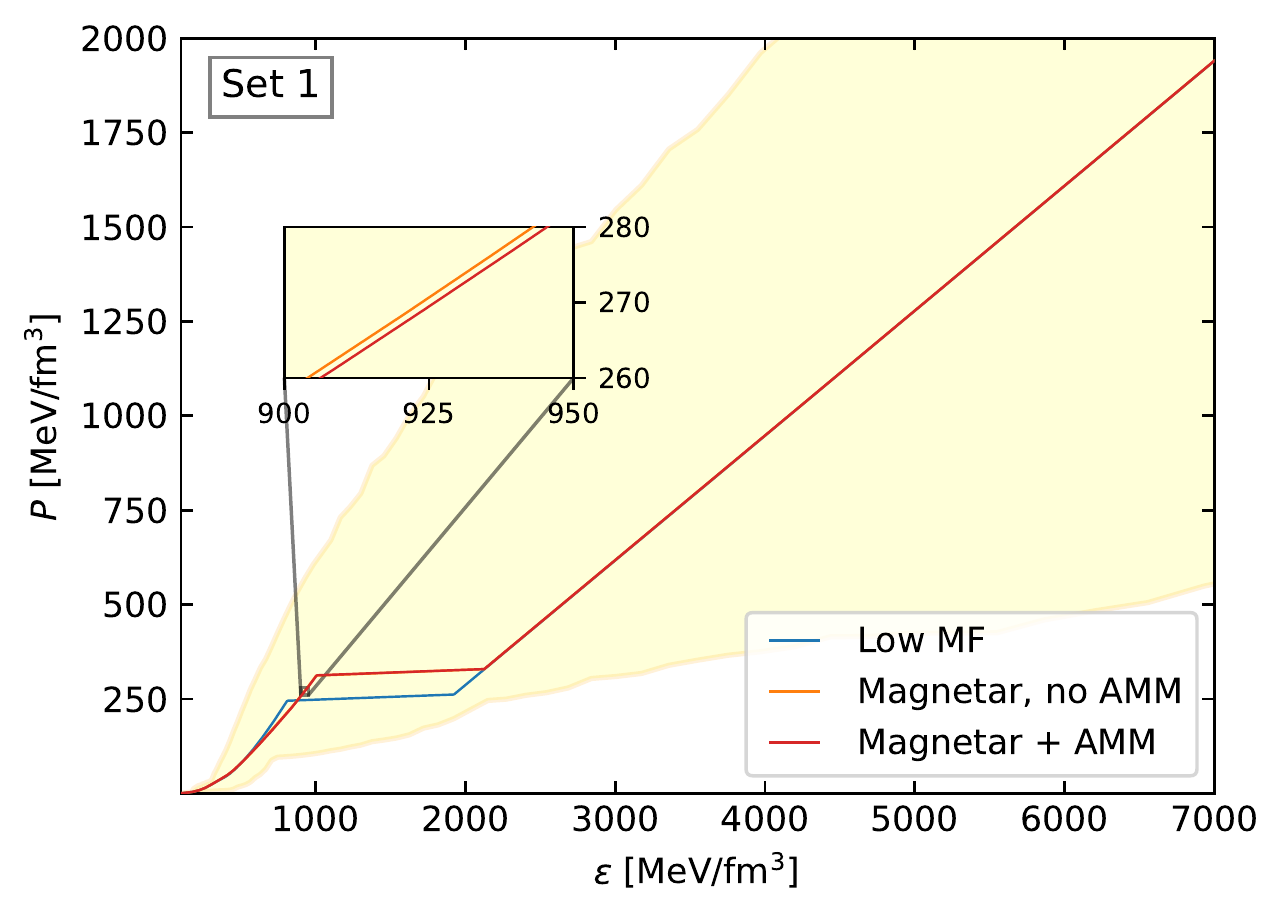}
    \includegraphics[width=0.45\linewidth,angle=0]{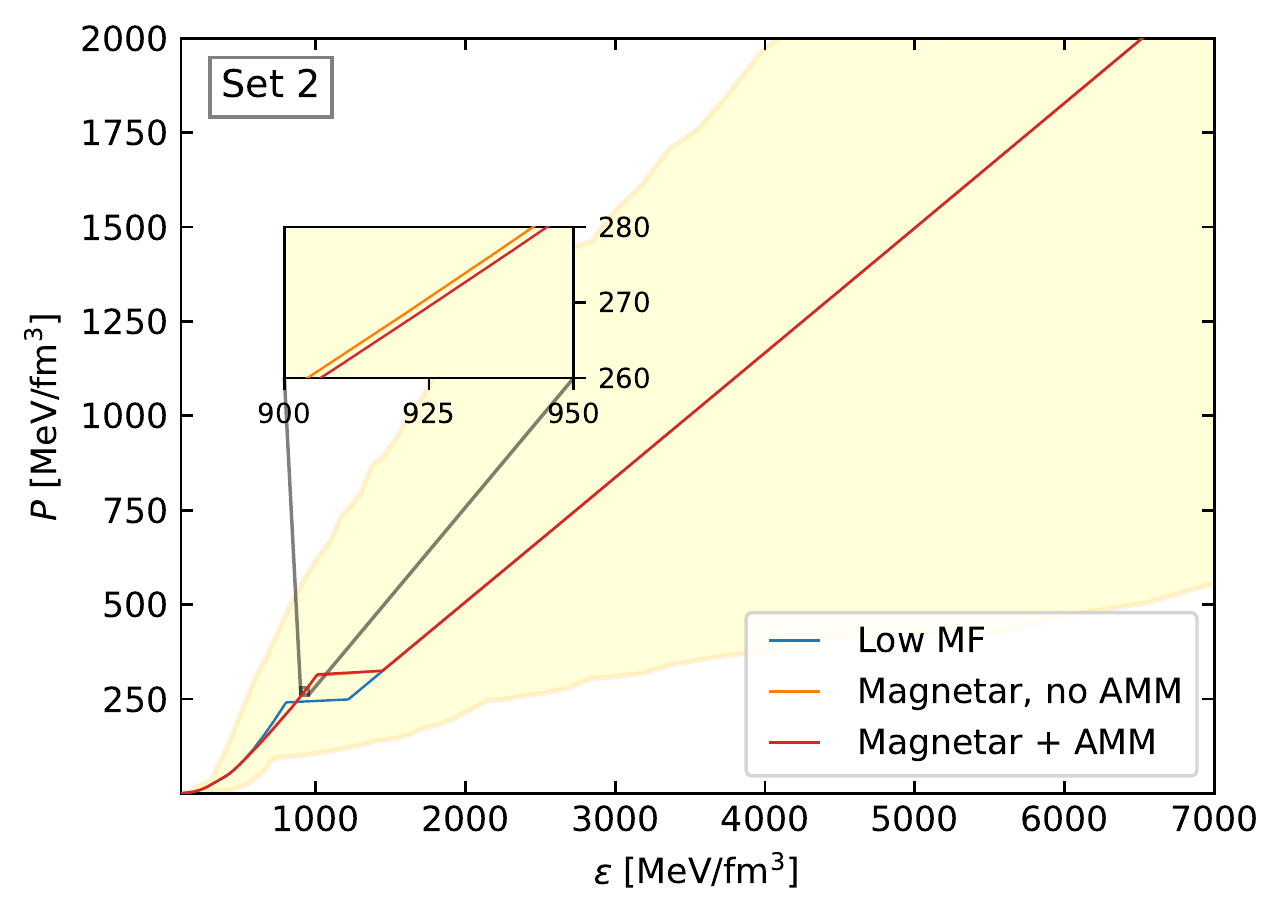}
    \includegraphics[width=0.45\linewidth,angle=0]{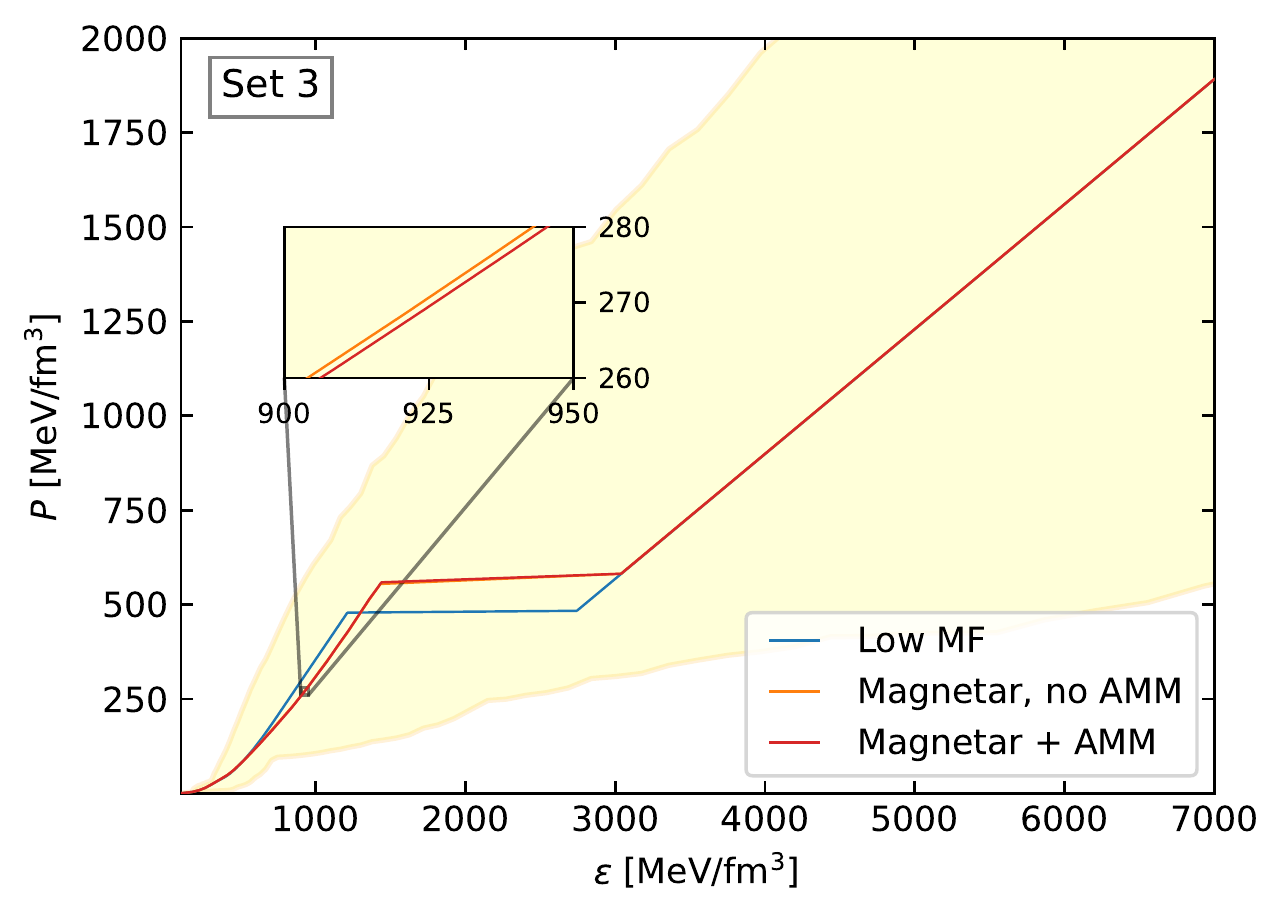}
     \caption{Hybrid magnetised EoSs for different MF scenarios and for the sets of Table~\ref{table:sets}. For all sets, increasing the MF strength softens the hadronic EoS and delays the onset of the phase transition, which occurs at higher pressures and energy densities. The enlarged boxes show the almost negligible effect of the inclusion of AMM in the energy spectrum of neutral particles. The yellow region indicates the nuclear \citep{Hebeler:2013nza} and hybrid interpolation \citep{Annala:2020efq} constraints.}
    \label{fig:eos}
\end{figure*}

In the sharp phase transition scenario, the nucleation time of the hadron-quark transition -the characteristic time during which a hadron (quark) fluid element is converted into quark (hadronic) matter- determines the behaviour of the matter in the neighbourhood of the hadron-quark interface inside the star \citep{Pereira:2018pte}. When a perturbation occurs in a HS, the fluid begins to oscillate with a characteristic timescale. In the neighbourhood of the hadron-quark interface, the fluid oscillates with this period and its pressure becomes alternatively higher and lower than the phase transition pressure, $P_t$. The probability that a fluid element under these conditions converts from hadron to quarks (or \emph{vice versa}) depends on the relation between the nucleation timescale and the oscillating timescale. If the nucleation timescale is much larger than the perturbation timescale, a fluid element oscillating around the transition surface will not undergo a phase transition and the interface position will oscillate along with the fluid; it is known as \emph{slow} conversion regime. On the contrary, within the \emph{rapid} conversion regime, when the nucleation timescale is much smaller than the oscillating timescale, the fluid elements convert almost immediately from one phase to the other, and the interface position remains stationary while the fluid oscillates \citep[for a more detailed discussion see][]{Pereira:2018pte,Mariani:2019mhs, lugonesuniverse7120493}.

As we will discuss later, these two, rapid and slow, extreme scenarios substantially modify the dynamical stability of HSs. As the hadron-quark nucleation timescale value is model dependent and uncertain, we will explore both possibilities and their astrophysical implications.

On the other hand, the MF introduces local anisotropies due to its field line geometry; parallel and perpendicular components of the matter pressure appear, $P_\parallel$ and $P_\perp$ respectively. Given the thermodynamic grand canonical potential, $\Omega$, these pressures can be written as
\begin{align}
P_\parallel &= - \Omega \, , \nonumber \\
P_\perp &= - \Omega - \mathcal{M}B \, ,
\label{pppper}
\end{align}
where $\mathcal{M}$ is the total matter magnetisation \citep{Blanford:1982mso,Gonzalez:2008msq},
\begin{equation}
    \mathcal{M}= - \partial \Omega / \partial B \rvert_{\mu_B}\, .
\label{magmatter}
\end{equation}
There also appears a pure electromagnetic contribution, $\propto B^2$, to the energy density and pressure. Hence, the energy-momentum tensor for the magnetised system reads \citep{Mariani:2019mhs}
\begin{align}
T_{\mu \nu} &=T^{\text{matter}}_{\mu \nu}+T^{\text{MF}}_{\mu \nu} \nonumber \\ 
&=\text{diag}(\epsilon+B^2/2, P_\perp+B^2/2, P_\perp+B^2/2,  P_\parallel-B^2/2) \, .
\end{align}
In the following subsection, we will study the effects of anisotropy and matter magnetisation on the EoS.

In this work we adopt the \emph{chaotic MF} approximation \citep{zel2014stars,Flores:2020gws}. Although locally we adopt a $z-$direction MF, globally it is an entangled combination of toroidal and poloidal contributions, and its field line directions change disorderly along the star interior. Within this hypothesis, the anisotropies compensate themselves from one spatial point to another and we can average the spatial components of the energy-momentum tensor. As the toroidal (poloidal) MF component tends to make the NSs prolate (oblate), in the chaotic scenario the global spherical symmetry is maintained and we obtain an effective isotropic pressure \citep{Bednarek:2003tio,Flores:2016pos, Mariani:2019mhs}:
\begin{equation}
	P=\frac{T_{1 1}+T_{2 2}+T_{3 3}}{3}=\frac{2 P_\perp+P_\parallel}{3} + \frac{B^2}{6}=P_{\text{matter}}+P_{\text{MF}} \, .
\label{pressure_prescription}	
\end{equation}

\subsection{Microphysics results}
\label{sub_micro}

In Fig.~\ref{fig:popH}, we show the hadronic particle population for different MF configurations, considering charge neutrality and $\beta$-equilibrium. In general, there are no significant differences in the particle fractions between the MF scenarios; only the leptons -electron and muons- and the $\Delta^0$ baryons show a noticeable change in cases of strong MF. Lepton fraction increases substantially and the electron population survives at densities of higher baryon number than in the case of a low MF, contrary to the muons, which fraction vanishes at lower densities. Particle population of $\Delta^0$ is strongly affected by AMM, which causes them to appear at lower densities. The inclusion of AMM has no significant influence on the rest of the particle populations. The fluctuations in the dotted and dashed curves are due to the filling of Landau levels, more visible in the case of leptons. The influence of strong MF in the EoS, when considering leptons, affects particularly the cooling of NS. Strong MF can activate the \emph{direct Urca} (dUrca) process in the core of these objects, which otherwise can only happen if the EoS modelling the star has a high fraction of protons. This is a fairly rigorous constraint for the EoS. The presence or absence of dUrca process is crucial for the cooling of NS due to neutrino emission \citep{Baiko:1999dup}. The earlier onset of $ \Delta^0$ in the magnetar+AMM scenario may also open up dUrca processes involving these particles, similar to what occurs in cases of $\Delta$'s nucleation \citep{Li:2018cbd, Raduta2021}.

In Fig.~\ref{fig:popQ}, we show the quark particle population for the two MF scenarios, considering charge neutrality and $\beta$-equilibrium. The continuous (dotted) lines indicate the low MF (magnetar) scenarios. As in the hadron phase, the strong MF significantly affects the lepton fraction. It can be seen that electrons are strongly affected by the strength of the MF. This effect is much more noticeable in the case of magnetars, where the dotted blue curve shows the characteristics oscillations due to Landau quantisation. The effect of the MF in the quark population is hardly noticeable. Muons fraction is negligible.

In Fig.~\ref{fig:popHyb}, we show the particle population of the hybrid EoS for the cases of Table~\ref{table:sets}, considering charge neutrality, $\beta$-equilibrium, and the Maxwell phase transition. The hatched grey area indicates the density jump due to the abrupt phase transition; the hadron phase corresponds to lower densities and, at higher densities, the quark matter phase occurs (almost at $\sim 5 n_0$ for Set~$1$ and $2$, and $\sim 6.25 n_0$ for Set~$3$). The MF strength does not affect the onset of the phase transition in terms of the $n_B$ value, but it acts on the particle population increasing the lepton fractions. Moreover, the inclusion of AMM favours the appearance of the $\Delta^0$ particle. The behaviour of the phase transition changes depending on the combination of the FCM parameters. This can affect both the onset of the phase transition and the width of the density jump, directly influencing the EoS, as we will show later, and the presence or absence of some hadrons, such as $\Xi^0$, $\Xi^-$ y $\Delta^0$.

In Fig.~\ref{fig:eos}, we show the resulting magnetised hybrid EoS for the cases of Table~\ref{table:sets}. As it can be seen, the effect of the different FCM sets is changing the onset and width of the phase transition. The pressure transitions corresponds to \mbox{$250~\textrm{Mev}/\textrm{fm}^3 \lesssim P_t \lesssim 500~\textrm{Mev}/\textrm{fm}^3$}, with energy density jumps varying in about $200~\textrm{Mev}/\textrm{fm}^3 \lesssim \Delta\epsilon \lesssim 2000~\textrm{Mev}/\textrm{fm}^3$. Increasing the MF softens the hadronic EoS, and the phase transition occurs at higher pressure and energy density for all the sets considered. The hybrid EoSs studied are within the yellow area, satisfying the nuclear \citep{Hebeler:2013nza} and hybrid interpolation \citep{Annala:2020efq} constraints. The enlarged boxes in Fig.~\ref{fig:eos} show the almost negligible effect of the AMM on the hybrid EoS. This implies that the influence of the AMM on the macroscopic properties of the stellar configurations like mass, radius and non-radial oscillations modes, becomes irrelevant. Therefore, from now on we will focus on only two scenarios: configurations with low MF and magnetars. 

\begin{figure}
    \centering
    \includegraphics[width=1\linewidth]{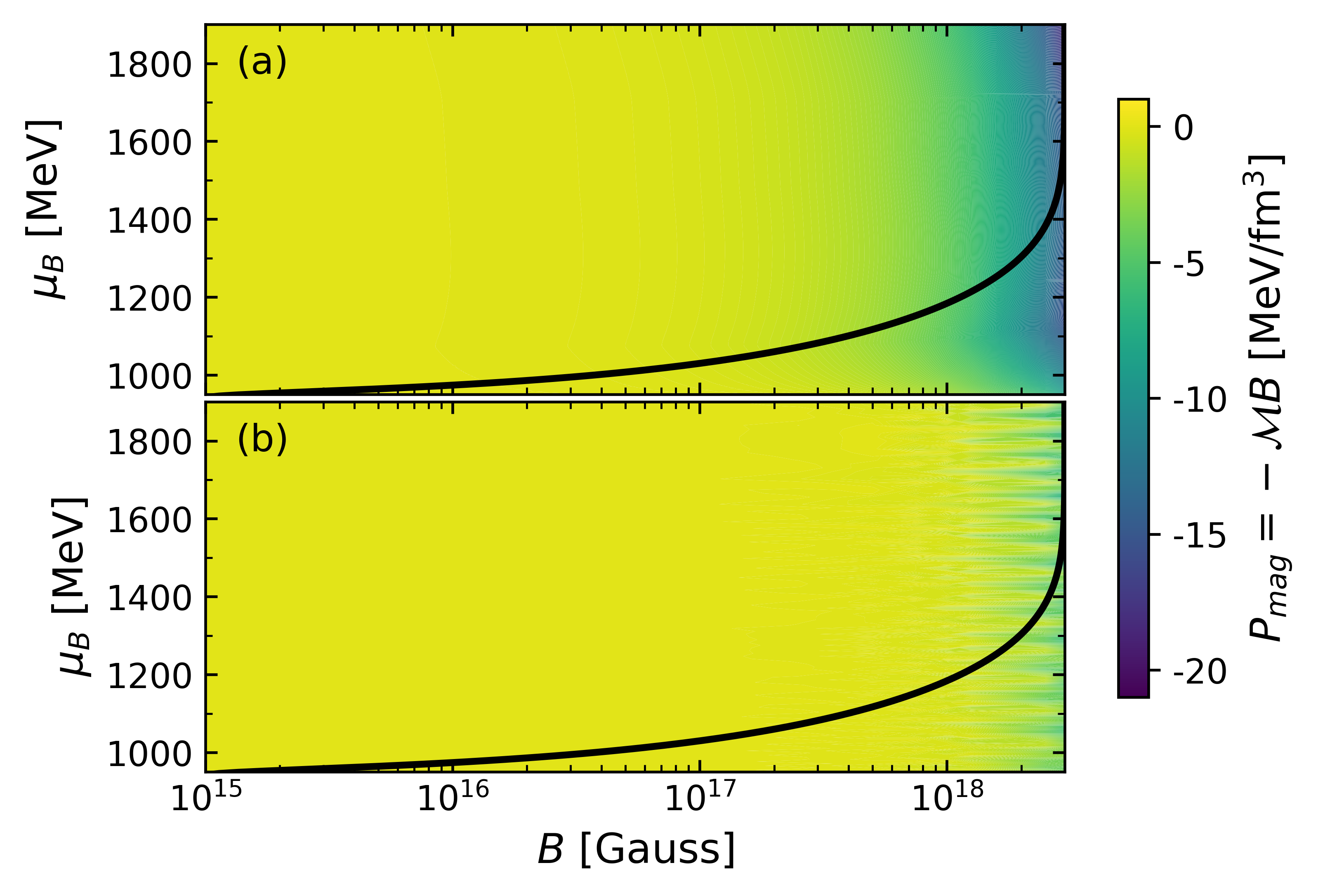}
    \caption{Magnetisation pressure, $P_{mag}$, in the $\mu_B$-$B$ plane. Panels (a) and (b) represent the hadron and quark phases, respectively. The black curves indicate the functional form of $B(n_B$), considering Eq.~(\ref{param}). In both panels, the magnetisation pressure becomes not negligible for MF above $\sim 2 \times 10^{18}$~Gauss. Therefore, $P_{mag}$ will be considered only in the quark phase.}
    \label{fig:mag}
\end{figure}

\begin{figure*}
    \centering
    \includegraphics[width=0.4\linewidth,angle=0]{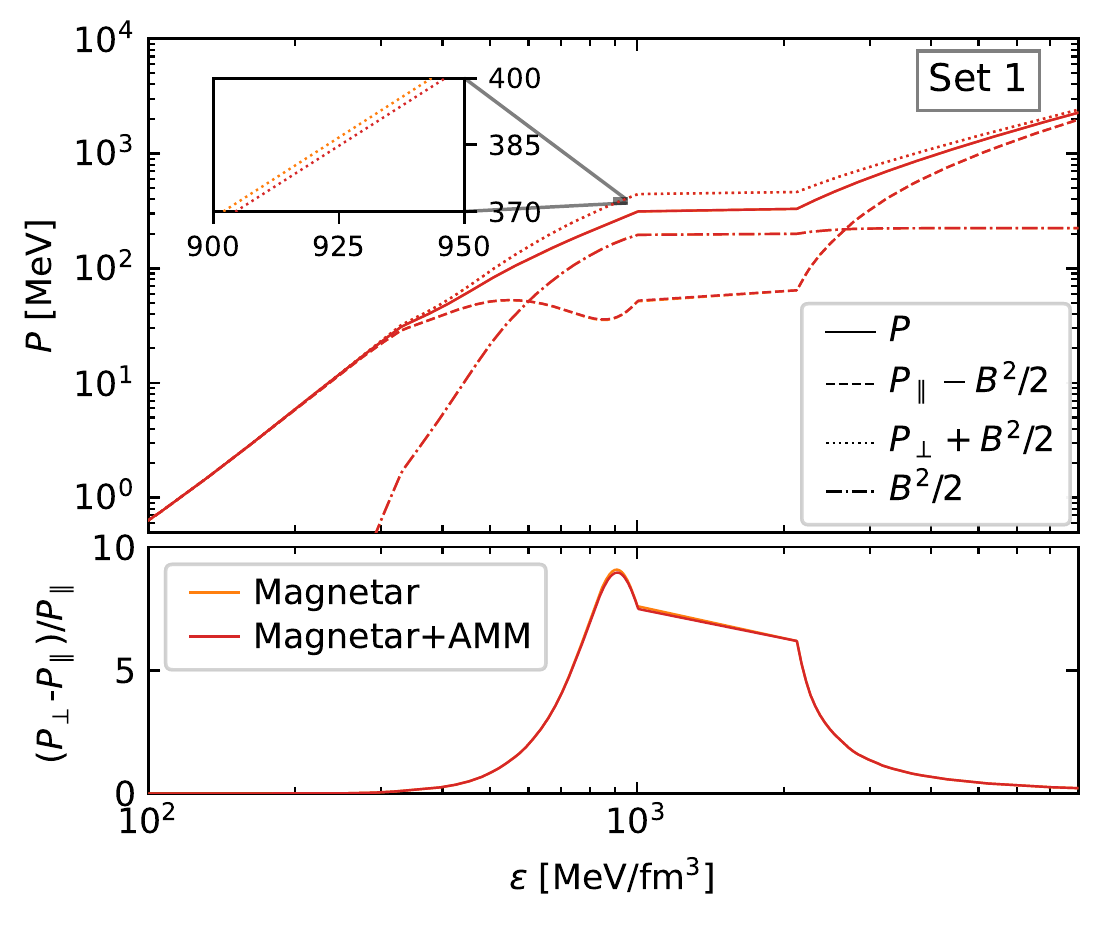}
    \includegraphics[width=0.4\linewidth,angle=0]{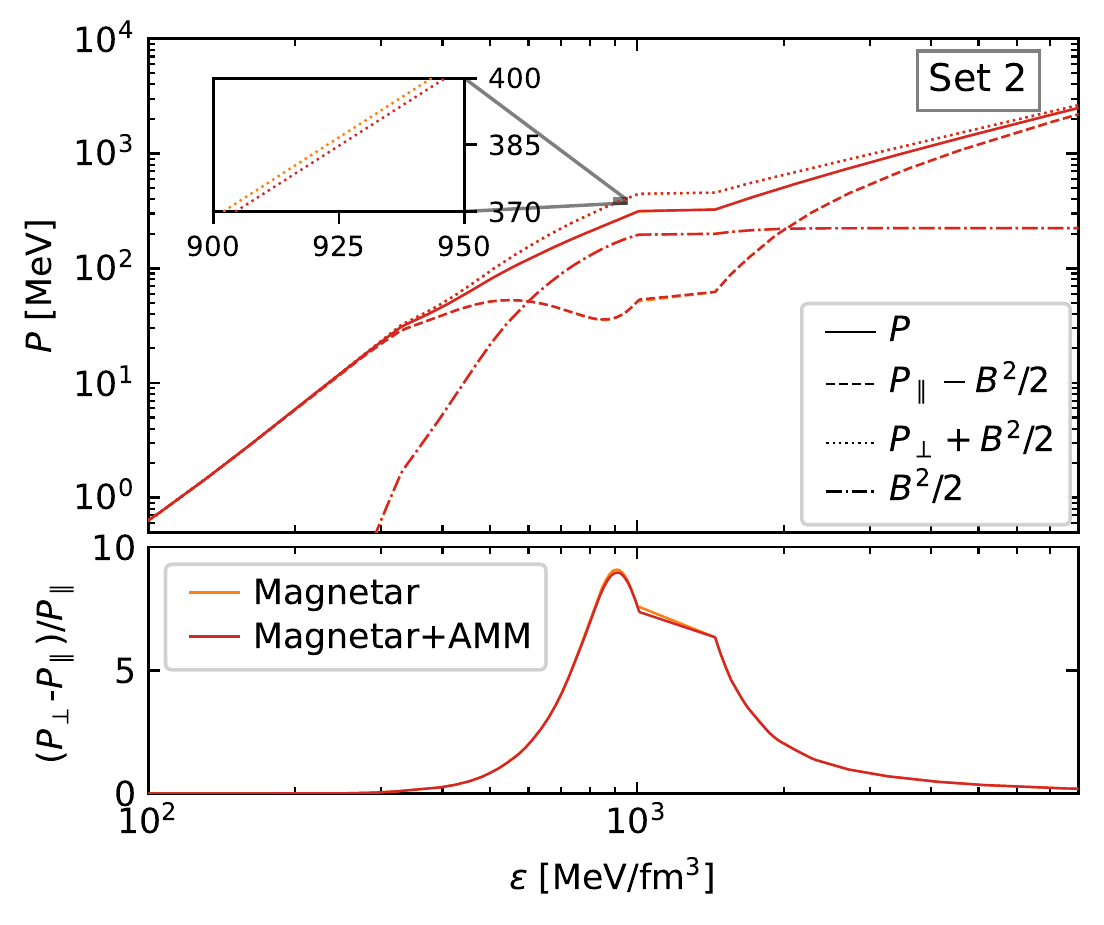}
    \includegraphics[width=0.4\linewidth,angle=0]{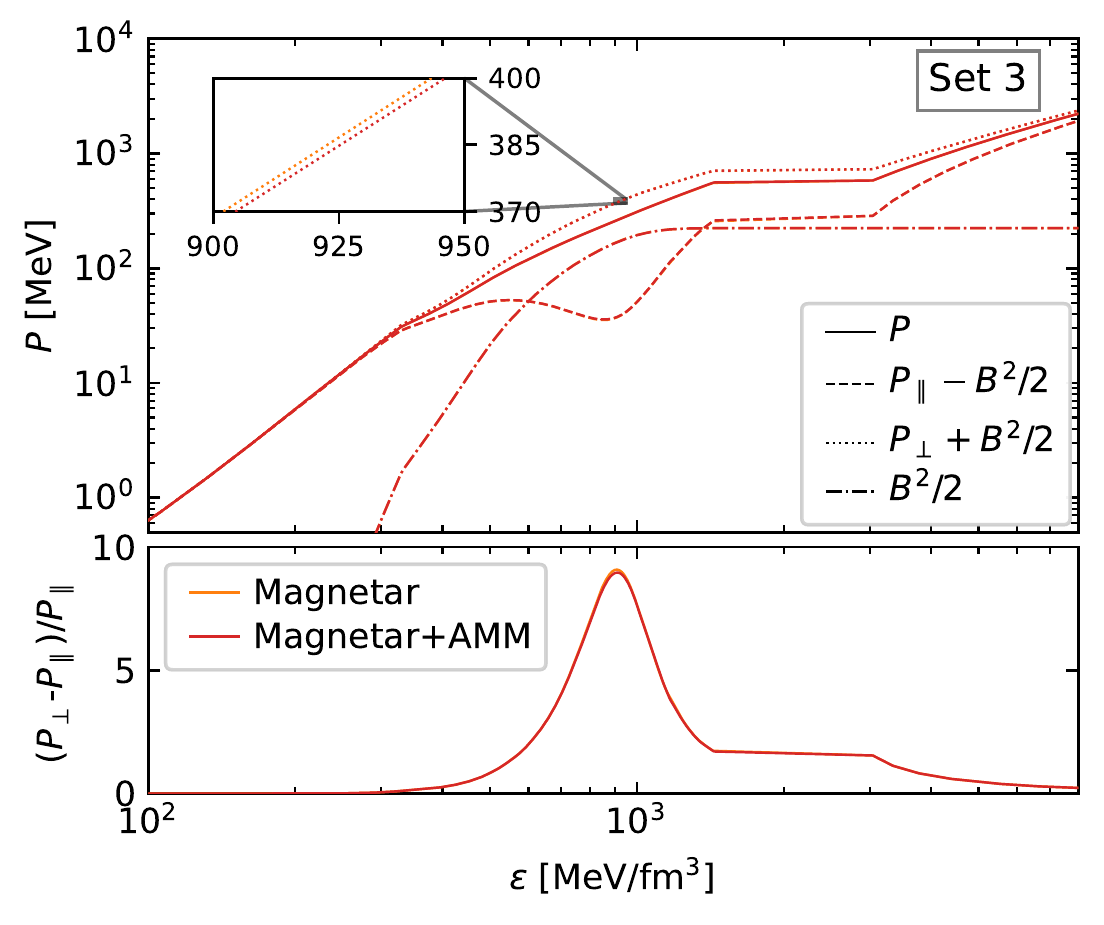}
    \caption{Upper panel: Pressure components of the hybrid magnetised EoS as a function of the energy density for the sets of Table~\ref{table:sets}, considering in detail the pressure anisotropy and the AMM contribution. Continuous line indicate the total pressure, $P$ -the effective isotropic pressure of Eq.~(\ref{pressure_prescription})-, composed by the parallel, $P_\parallel-B^2/2$, and perpendicular, $P_\perp+B^2/2$ components (dashed and dotted lines, respectively). The pure magnetic term, $P_B = \pm B^2/2$, (presented in absolute value and with dashed dotted line) produces the main difference between $P_\parallel$ and $P_\perp$. Due to the parametric function of Eq.~(\ref{param}), the MF tends to a constant value as the energy density increases, and this difference between the pressure components vanishes for high enough densities. The enlarged box shows the negligible effect of the AMM for $P_\perp$, but the same happens with $P_\parallel$. Lower panel: the relative difference between $P_\parallel$ and $P_\perp$ increases with the increasing of the MF strength, reaching a peak value in the hadron phase before the phase transition, and progressively vanishing in the quark phase.}
    \label{fig:presiones}
\end{figure*}

In Fig.~\ref{fig:mag}, we show one of the quantities contributing to the anisotropy of the pressure, the magnetisation pressure, $P_{\rm mag}$, for hadron -panel (a)- and quark -panel (b)- phases. We present $P_{\rm mag}$ as a function of both the baryon chemical potential, $\mu_B$, and the MF, $B$. In this plane, we also show the MF parametrization function of Eq.~(\ref{param}), indicated with a black curve. Our results show that the magnetisation pressure only is not negligible for MF values, $B \gtrsim 2 \times 10^{18}$~Gauss. Thus, $P_{\rm mag}$ will be only considered for the quark phase, as such intensity of MF strength values is expected to exist only in the inner core of HSs.

Finally, in Fig.~\ref{fig:presiones}, we show the effective isotropic pressure of Eq.~(\ref{pressure_prescription}), $P$, and the parallel, $P_\parallel-B^2/2$, and perpendicular, $P_\perp+B^2/2$, pressures, indicated by continuous, dashed, and dashed dotted lines, respectively, considering the presence and absence of AMM contribution. The low MF scenario does not present any anisotropy or significant pure magnetic contribution from the pressure, $P_B = \pm B^2/2$ (dashed dotted line). The anisotropy because of the splitting of $P_\parallel-B^2/2$ and $P_\perp+B^2/2$ is noticeable mainly in the hadron phase, before the phase transition, where the contribution of $P_B$ is dominant. As the energy density increases, and due to the functional form of the MF given in Eq.~(\ref{param}), $P_B$  becomes a constant value. Hence, the relative difference between $P_\parallel-B^2/2$ and $P_\perp+B^2/2$ reaches a maximum in the hadron phase, then it begins to diminish and, for a high enough density in the quark phase, it vanishes. The negligible contribution of the AMM, as has already been shown in Fig.~\ref{fig:eos}, occurs in both components of the pressure (showed in the enlarged box only for $P_\perp+B^2/2$, for simplicity). As has been previously discussed, we adopt the chaotic MF approximation, where the anisotropy of the pressure obtained does not represent a global macroscopic property of the stellar configurations studied in this work, but it is considered to be only local.

\section{Stellar Structure and Stability}
\label{structure}

Using the chaotic MF approximation, Eq.~(\ref{pressure_prescription}), to construct the hybrid magnetised EoS, $P(\epsilon)$, we integrate the stable stellar configurations through the spherically symmetric Tolman-Oppenheimer-Volkoff (TOV) structure equations to obtain the gravitational mass and radius of a given stellar configuration.

Besides mass and radius, another relevant quantity is the dimensionless tidal deformability, $\Lambda$, constrained by the GW observatories. The details of the $\Lambda$ calculation for HSs with abrupt phase transitions are presented in \citet{Han:2018tdw}.

The solutions of TOV equations provide stellar equilibrium configurations regardless of their stability. The stability study implies analysing the response of equilibrium configurations to small radial disturbances and inspecting the values of the fundamental radial eigenfrequency, $\omega_{0}$. The analysis of the stability of compact objects, in the framework of General Relativity, started in the decade of 1960, where key results related to the stability of spherically-symmetric cold-catalysed one-phase compact stars were obtained. The traditional stability criterion states that under this scenario, if stellar configurations are stable, then $\partial M/ \partial \epsilon_c > 0$ and they are unstable in regions where $\partial M/\partial \epsilon_c < 0$ \citep{Chandrasekhar:1964dio,Harrison:1965gta,Shapiro:1986bhw}. As we already mentioned, this picture can change drastically when sharp hadron-quark phase transitions are taken into account. The key ingredient, in this case, is the speed of the hadron-quark conversion, compared to the oscillation period of a fluid element at the hadron-quark interface in the star. In the work by \citet{Pereira:2018pte}, it has been shown that two limiting cases are interesting to study: slow and rapid conversions. The different nature of this physical process has deep implications on the boundary conditions imposed at the hadron-quark interface, when solving the radial oscillation eigenvalue problem for HSs with a sharp discontinuity in the energy density profile inside the star.
    
If the hadron-quark conversion is rapid then the standard equilibrium criterion is conserved. However, if slow conversion is taken into account, the fundamental radial eigenfrequency might not change its nature (from stable to unstable) at the critical points of $M(\epsilon_c)$ relationship. For this reason, an extended stability branch exists despite being a region where $\partial M/\partial \epsilon_c < 0$, as the fundamental radial mode might not become imaginary at the critical points of a Mass-Radius diagram \citep[to see this approach in different astrophysical scenarios see, for example, ][]{Tonetto:2020dgm,Rodriguez:2021hsw,Curin:2021hsw}.

\subsection{Results for Stellar Structure and Stability}

As we mentioned in the Subsection~\ref{sub_micro}, due to the negligible effect of the AMM on the hybrid EoS, we will present the astrophysical results for only two scenarios: low MF and magnetar.

\begin{figure}
    \centering
    \includegraphics[width=0.9\linewidth,angle=0]{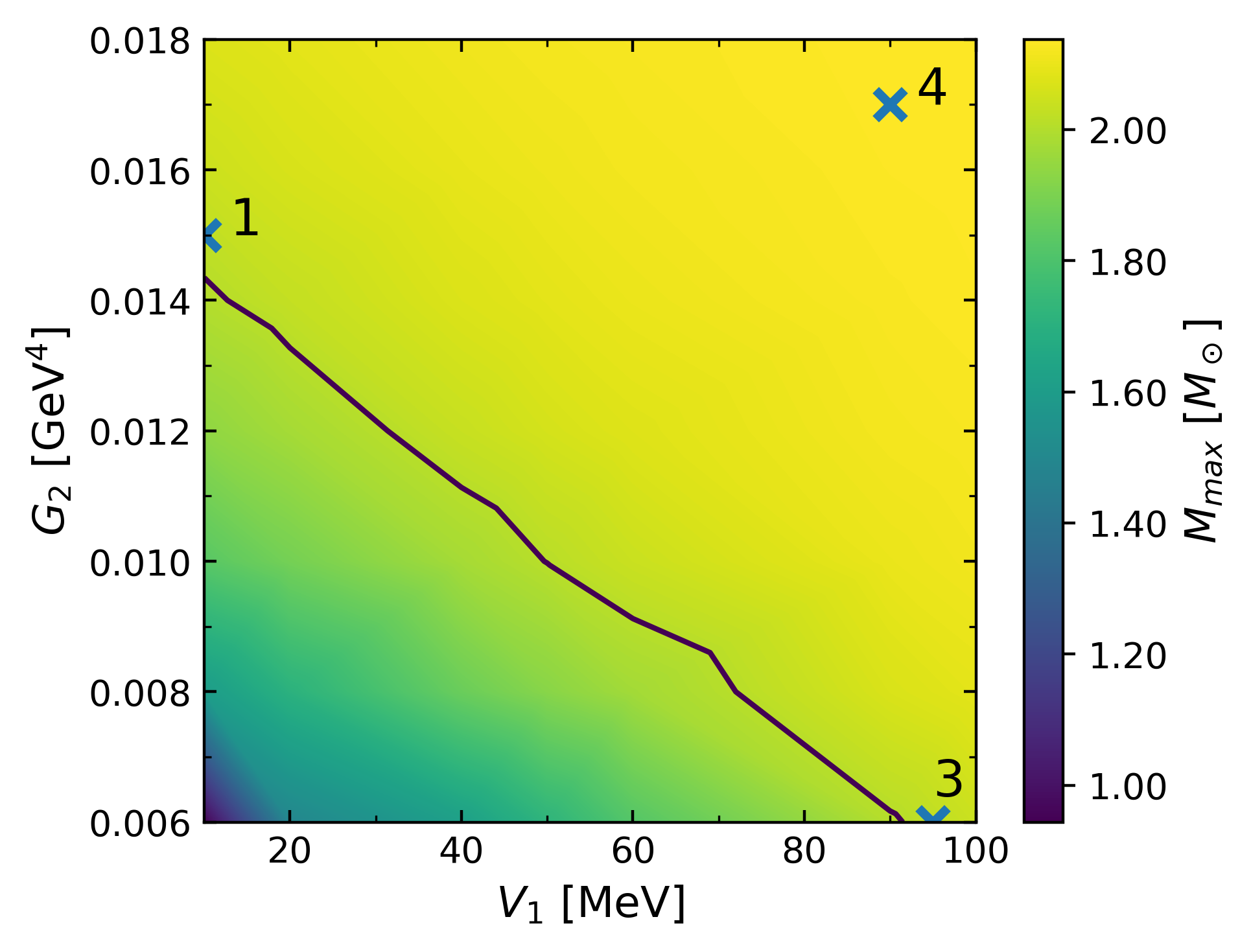}
    \caption{Maximum mass for each family of stars in the $G_2$-$V_1$ plane for the low MF scenario. The $2.01~M_\odot$ level is represented with the black curve. The numbered crosses indicate the representative selected sets chosen to present the astrophysical results (see Table~\ref{table:sets} for details). }
    \label{fig:mmax}
\end{figure}

We integrated the TOV equations for a wide range of the FCM parameters to map the combinations of $G_2$ and $V_1$ satisfying the $2.01~M_\odot$ constraint for the maximum mass configuration. This procedure was done considering only the low MF scenario, since the mass constraints are not expected to come from magnetar-type objects. In Fig.~\ref{fig:mmax}, we display the $G_2$-$V_1$-$M_{max}$ relationship through a colour map. The black curve indicates the $M_{max}=2.01~M_\odot$ level. As it can be seen, the maximum mass increases for higher values of $V_1$ and/or $G_2$ combination; the reason of this increasing is that the hadron-quark phase transition is pushed to higher pressure values. Moreover, when imposing a minimum mass for the compact object's family consistent with the $2~M_\odot$ pulsars, the appearance of the quark core produces the masses of the stellar family to stop increasing, and so, all the hybrid curves have a lower maximum mass than the pure hadronic curve. In this sense, within our model, the pure hadronic EoS imposes an upper limit for the possible maximum mass value, \mbox{$M^{\textrm{had}}_{max} \sim 2.2~M_\odot$}, and, as $V_1$ and/or $G_2$ increases, the maximum mass of the hybrid family asymptotically approaches this value.

In the $G_2$-$V_1$ plane, we also mark the combination values of the three representative sets, ($V_1$, $G_2$), we select to construct HSs. These sets are qualitatively representative of the general behaviour of our model, and are detailed in Table~\ref{table:sets}.

\begin{table}
\centering
\begin{tabular}{|c|c|c|}
\hline 
Set \# & $V_1$ [MeV] & $G_2$ [GeV$^4$]  \\ \hline
1 & 10 & 0.015 \\
2 & 95 & 0.006 \\
3 & 90 & 0.017 \\
\hline
\end{tabular}
\caption{Representative selected sets of the FCM parameters, used for the quark phase. The three sets satisfy the current observational and are qualitatively representative of the general behaviour of our model.}
\label{table:sets}
\end{table}

\begin{figure*}
    \centering
    \includegraphics[width=0.45\linewidth,angle=0]{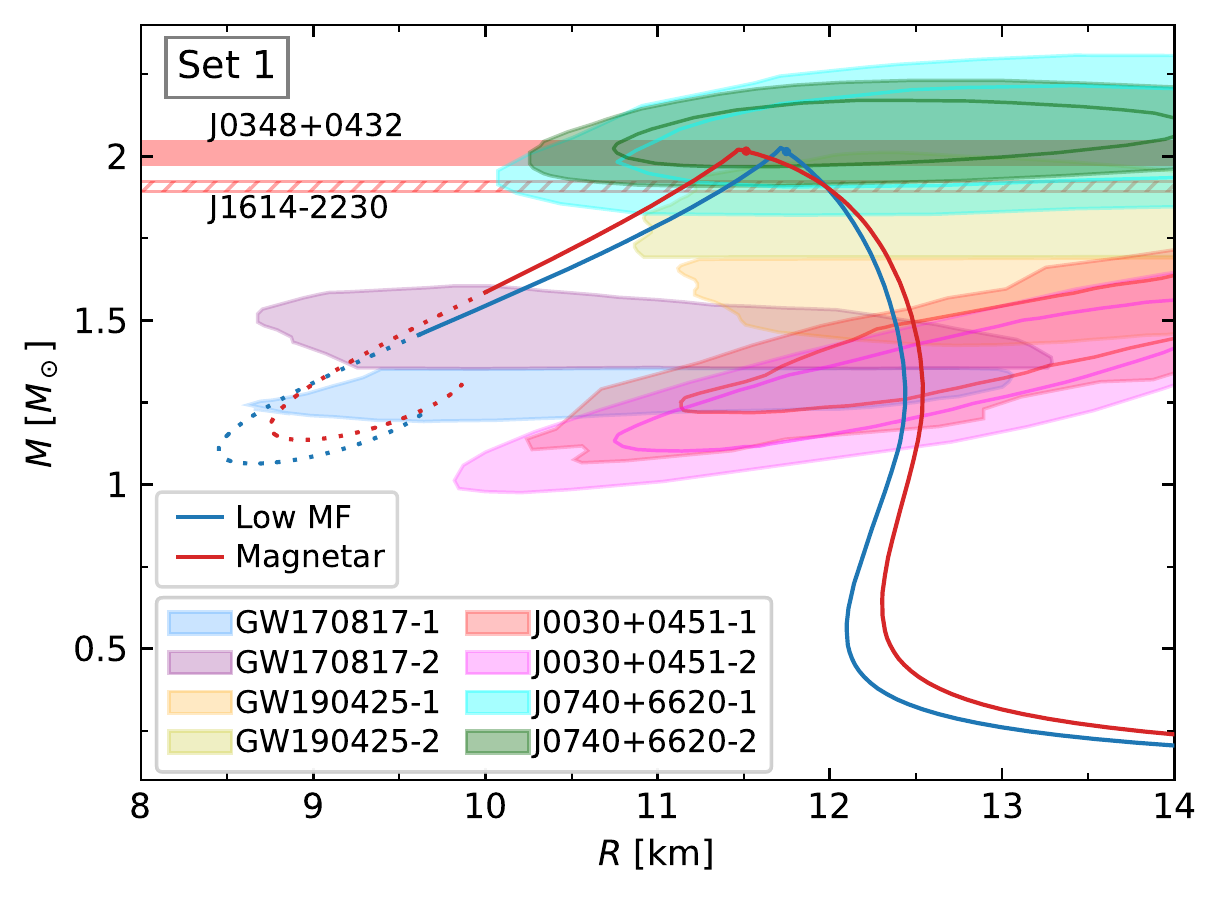}
    \includegraphics[width=0.45\linewidth,angle=0]{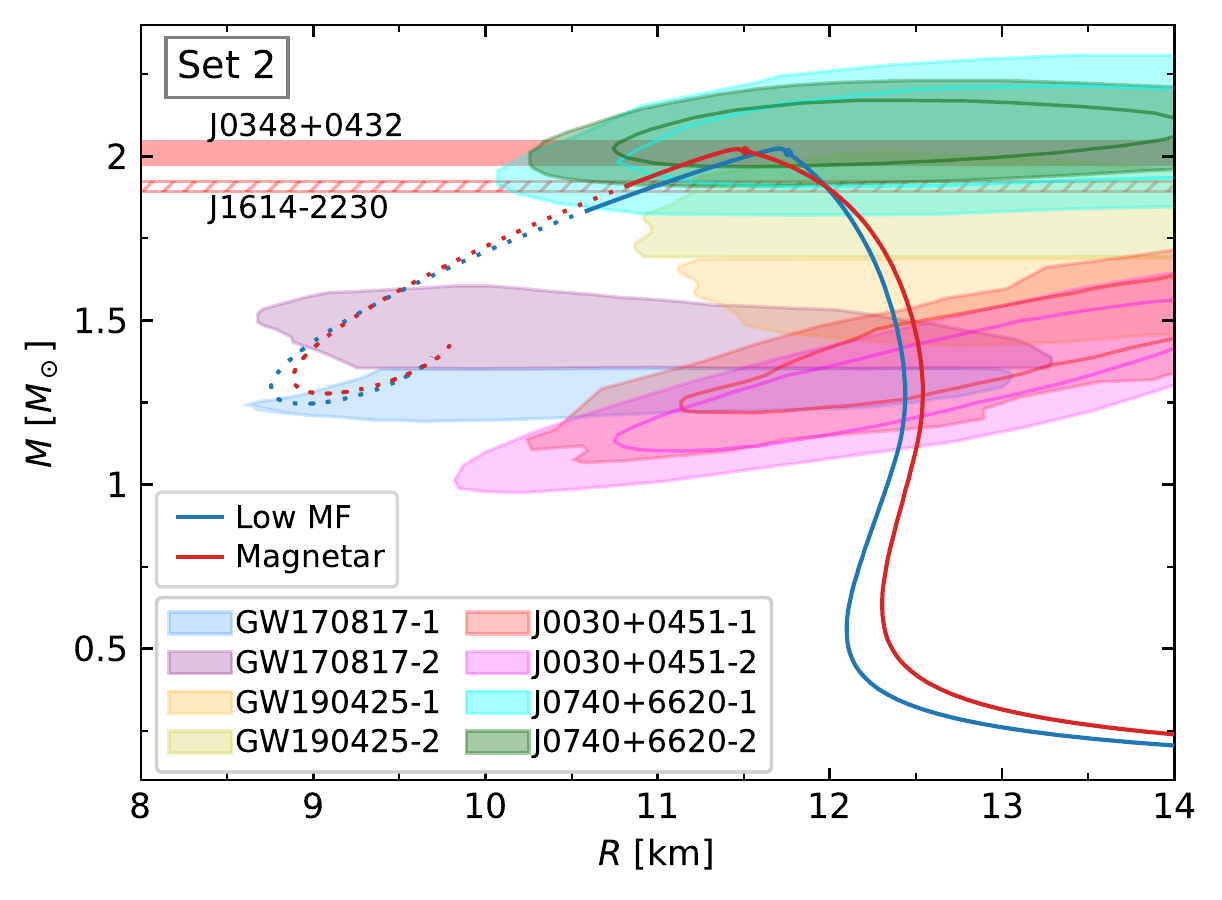}
    \includegraphics[width=0.45\linewidth,angle=0]{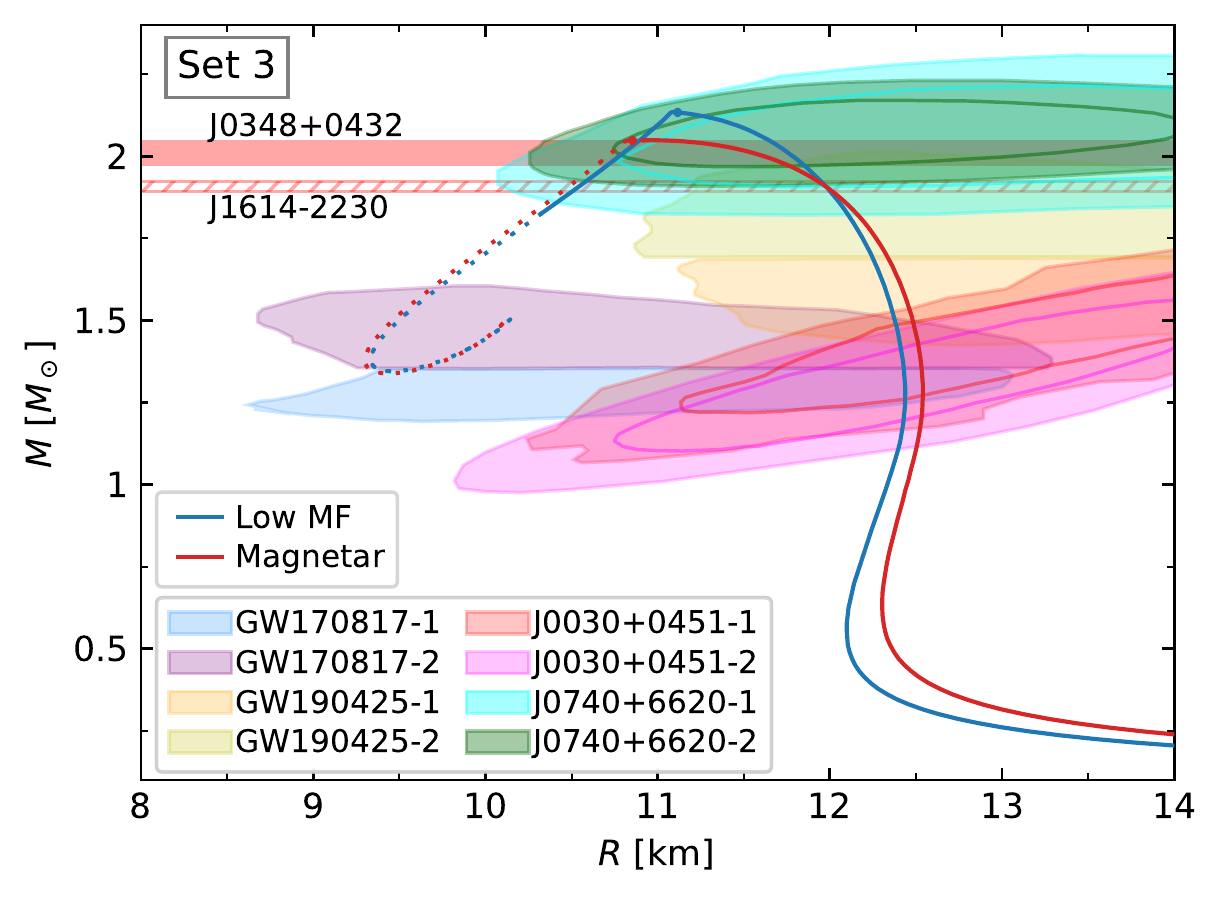}
    \caption{Mass-radius relationships for the sets of Table~\ref{table:sets}. When slow conversions are assumed, all points of the continuous curves are stable configurations; if conversions are rapid the stability is preserved only up to the maximum mass configurations. In both scenarios the dotted curves represent unstable stars. The rounded dot over each curve indicates the appearance of the quark matter core. All the curves have a purely hadronic branch and a hybrid slow conversion branch except for the magnetar case of Set~$3$. In this particular case, quark matter appear after the maximum mass configuration and therefore, the extended stability branch is absent, and this stars' family is only hadronic. For the Set~$1$, the slow extended stability branch reach the high-mass component of the GW170817 merger event, suggesting the possibility of a slow HS with a purely hadronic NS merger. The effect of the MF not always produces a higher maximum mass than for the low MF curve, as it can be seen from Set~$3$ (see details in text). We also present astrophysical constraints from the \mbox{$\sim 2~M_\odot$} pulsars \citep{Arzoumanian:2018tny, Fonseca:2021rfa}, GW170817 \citep{Abbott:2017oog} and GW190425 \citep{Abbott:2020goo} events, and NICER observations \citep{Miller:2019pjm,Riley2019anv,Riley:2021anv, Miller:2021tro}.}
    \label{fig:mraio}
\end{figure*}

For the representative sets of Table~\ref{table:sets}, we analyse the astrophysical results in detail. In Fig.~\ref{fig:mraio}, we show the mass-radius ($M$-$R$) plane containing the obtained stellar configurations and all the current astrophysical constraints. For all the curves, the rounded dot marks the onset of the quark matter core; from this dot towards larger radii the configurations are purely hadronic and toward smaller radii are hybrid. Further, the point where the stability is lost depends on the phase conversion scenario, slow or rapid. As follows from the radial perturbations stability analysis, in the rapid scenario the traditional stability criteria is still valid: the stability is preserved from larger radii up to the maximum mass configurations. Hence, within this stability regime, the possibility to obtain HSs is only marginal, since the quark matter core appears just before the maximum mass. In contrast, in the slow conversion scenario, stability is preserved along the continuous curve up to the last stable \emph{terminal mass} configuration, where the dotted curve begins. The slow extended stability branch, since it only exists in discontinuous EoSs cases, is always hybrid. The length of the extended stability branch also depends on the FCM parameters; in general, as $V_1$ or $G_2$ increases, the extended branch shortens. In particular, Set~$1$ shows the longest slow extended stability branch that could explain the stellar high-mass component of the GW170817 binary system. Set~$3$ panel shows that, in the magnetar curve, terminal mass coincides with maximum mass and the slow extended stability branch is absent. This happens since, in this case, the phase transition occurs after the maximum mass configuration. In all cases, the $M$-$R$ relationships satisfy the astrophysical constraints through the classic-stability hadronic branch.

For a given hadronic and quark EoS, the effects of the magnetised crust EoS used in this work \citep{Mutafchieva:2019rol} is to shrink the radius, particularly for low mass stars, e.g. for $M \simeq 0.75 M_\odot$, $\Delta R \simeq 0.05$~km, if we compare with the results considering the magnetised crust EoS of \citet{Lai:1991ceo}, used in \cite{Mariani:2019mhs}. As the mass increases, the difference in radius between the two crusts decreases , e.g. for $M \simeq 2.02 M_\odot$, $\Delta R \simeq 0.01$~km. The mass value of the maximum mass configuration is the same considering either of the two crusts.

Regarding the effect of the strong MF, we find that not always the magnetar models have a higher maximum mass than the low MF models, unlike other works \citep{Rabhi:2009qhp,Mariani:2019mhs,Thapa:2020eos,Rather:2021hmn}; in our model, Sets~$1$ and $2$ show slightly higher maximum masses for magnetars, but for Set~$3$ the low MF family has a higher maximum mass configuration than magnetars. The work of \citet{Flores:2020gws} also shows that dependence with the EoS model, the higher maximum mass alternatively corresponds to a low MF or magnetar scenario. Considering the EoSs results of Fig.~\ref{fig:eos}, the difference between the pressure transition value of Sets~$1$/$2$ and Set~$3$ seems to be the cause of this behaviour. Also, after the maximum mass configurations, the magnetar extended stability branches are always shorter than the corresponding low MF branches; Set~$3$ shows the extreme case where the extended stable branch does not exist in the magnetar curve. In all cases, the magnetar extended branches have higher mass than the low MF extended branches, for a given radius.

It is interesting to note that, regardless of the set considered, the intersection between the low MF and the magnetar curves occurs in the same region of the $M$-$R$ plane, $M \simeq 1.9 M_\odot$ and $R \simeq 12.5$~km. For configurations with lower masses than this value, magnetars have larger radii than the low MF case; magnetars have smaller radii for higher masses.

\begin{figure*}
    \centering
    \includegraphics[width=0.45\linewidth,angle=0]{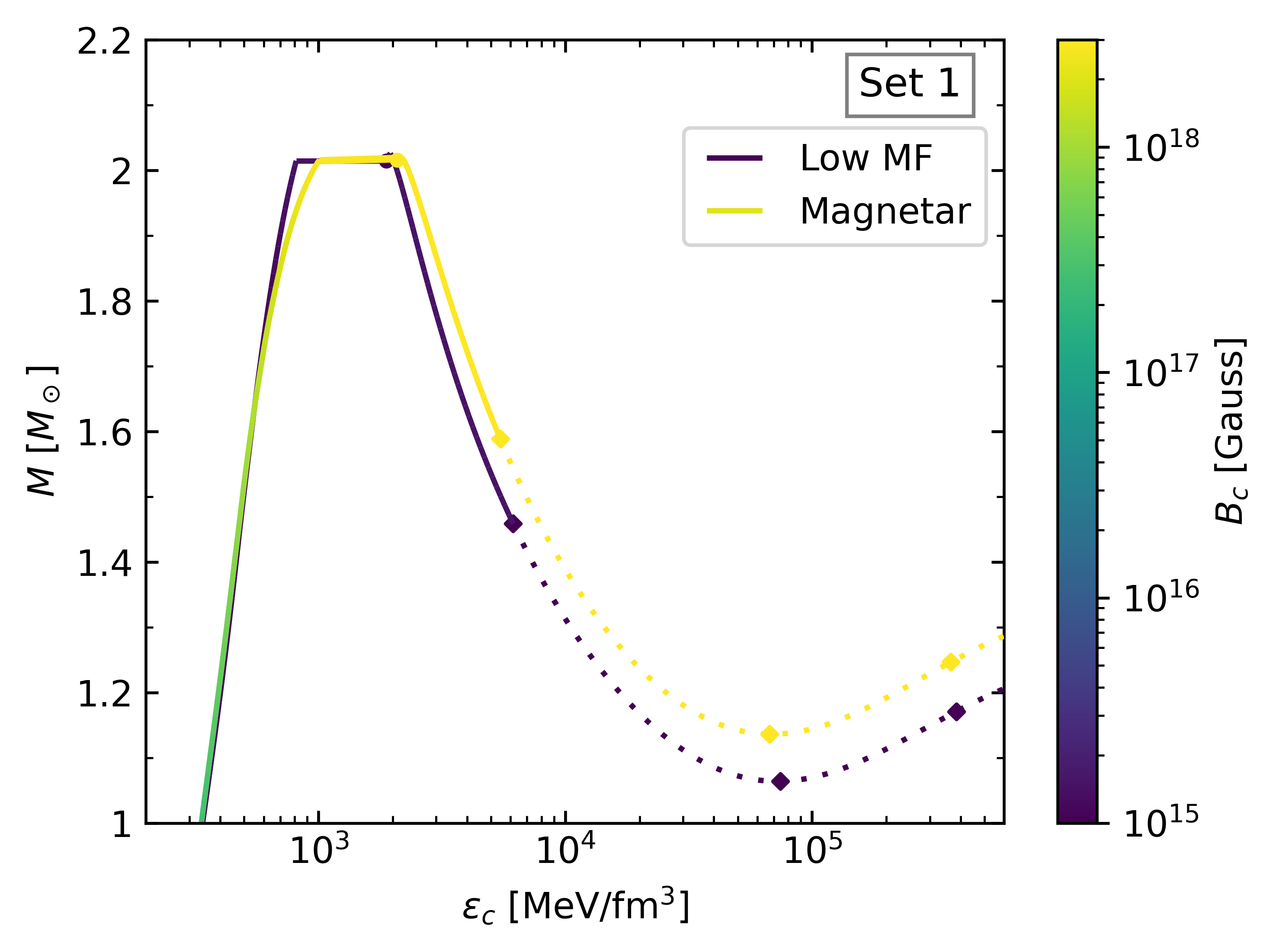}
    \includegraphics[width=0.45\linewidth,angle=0]{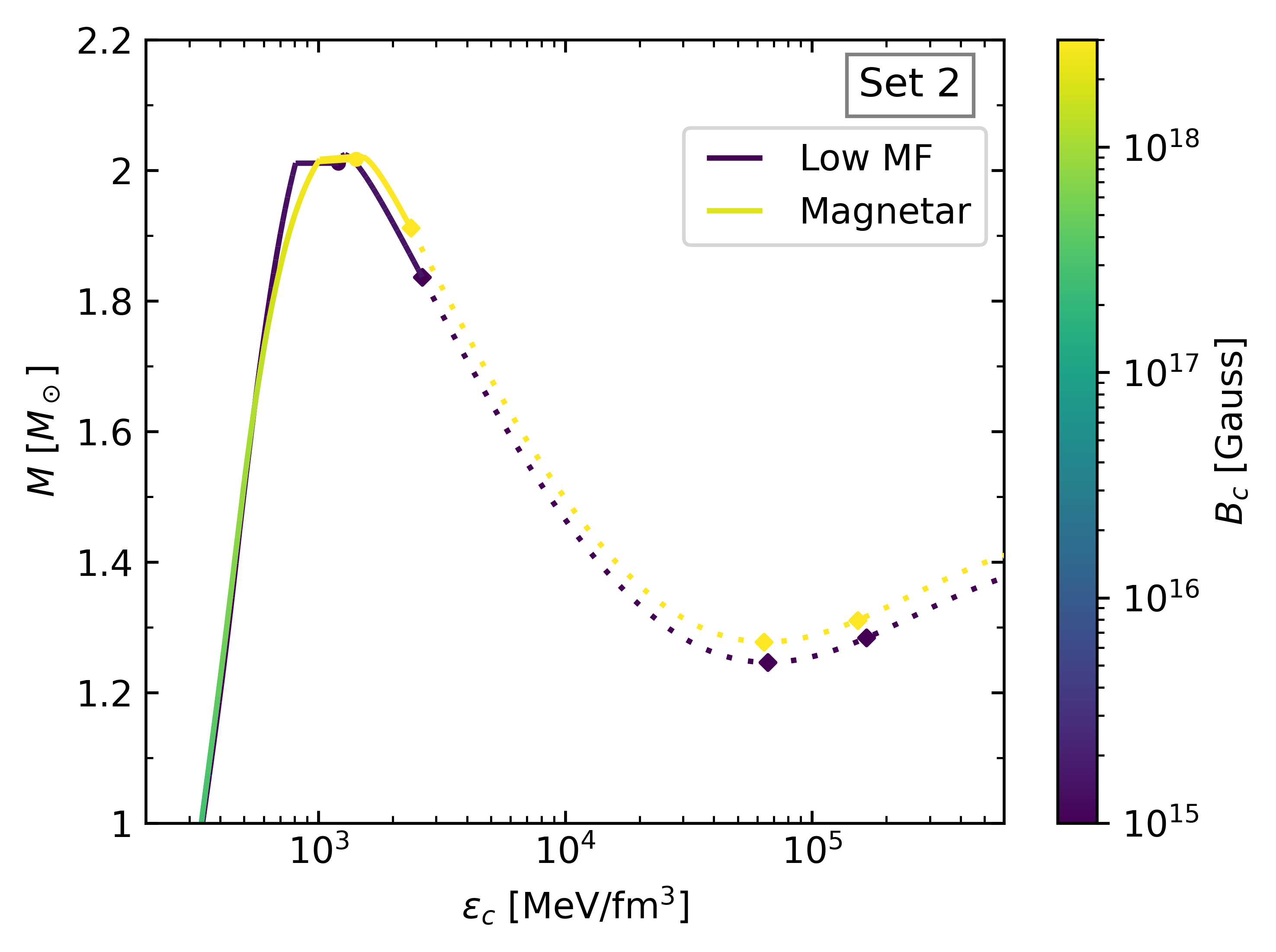}
    \includegraphics[width=0.45\linewidth,angle=0]{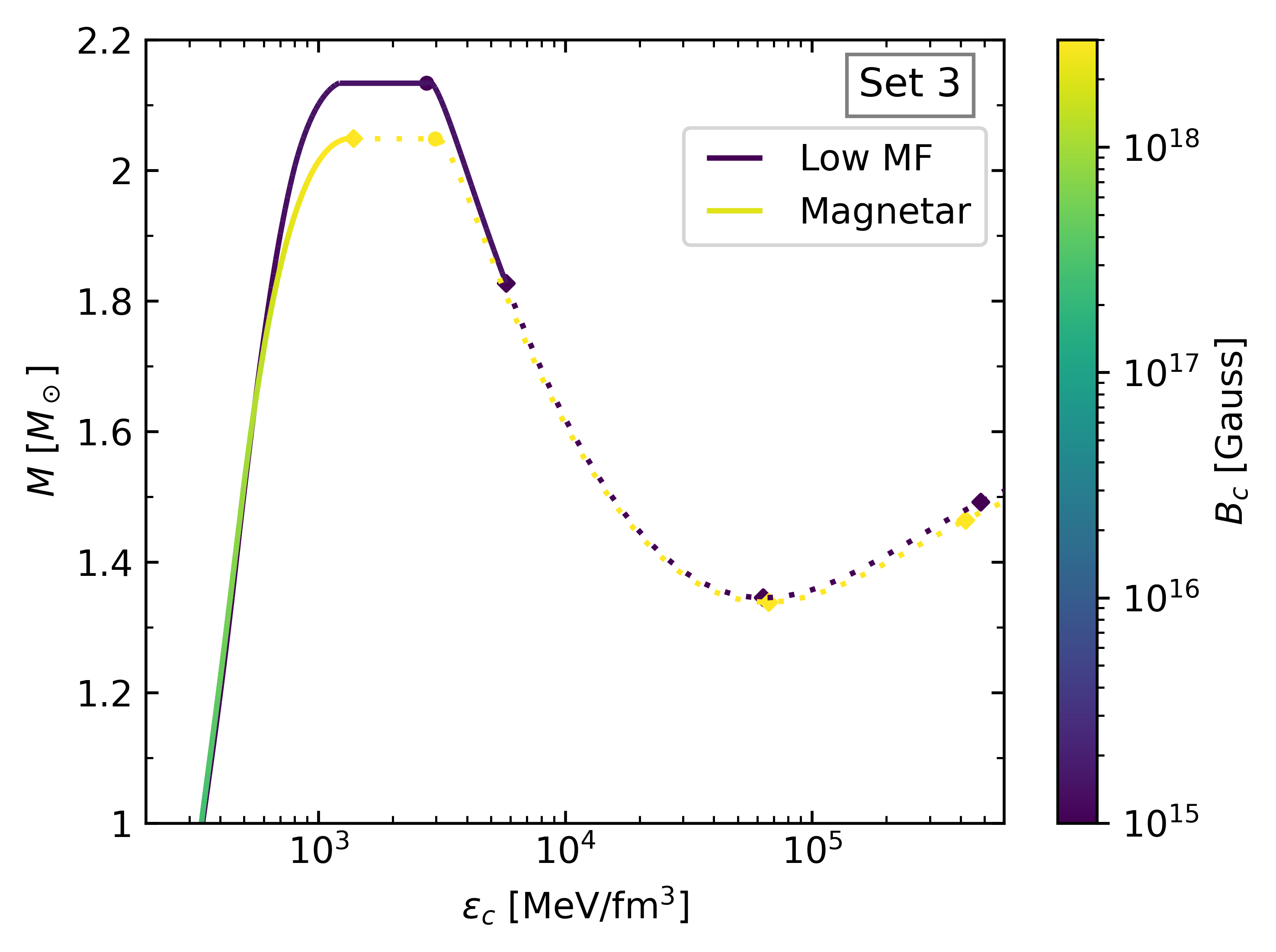}
    \caption{Mass-central energy density relationship for the representative sets of Table~\ref{table:sets}. The colour of each curve indicates the central MF, $B_c$, for each star. In the rapid scenario, the stability is preserved up to the maximum mass configuration; in the slow scenario, the stability is preserved up to the terminal mass, indicated with the first diamond dot separating the continuous from the dotted curves. The following diamond dots indicate the change in the sign of the successive squared eigenfrequencies, ($\omega_1^2$, $\omega_2^2$) within the slow conversion scenario. As it can be seen, these points does not necessarily coincide with the critical points of the curves, as it occurs in the rapid conversion scenario, and the $\partial M / \partial \epsilon_c < 0$ condition is no longer sufficient nor necessary to identify unstable configurations (see details in the text).}
    \label{fig:meps}
\end{figure*}

In Fig.~\ref{fig:meps}, we show the mass-central energy density ($M$-$\epsilon_c$) relationship for the sets of Table~\ref{table:sets}. The colour of each line indicates the value of the central MF, $B_c$, for each stellar configuration. As in Fig.~\ref{fig:mraio}, in the rapid scenario the stability is preserved only up to the maximum mass configuration, but in the slow scenario the stability remains up to the terminal mass, indicated with the first diamond dot that separates the continuous from the dotted curves. At this point, the fundamental radial mode frequency, $\omega_0$, vanishes. Hence, for higher central densities, $\omega_{0}^2 < 0$ and the configurations are unstable. The following diamond dots indicate where the successive radial frequency modes ($\omega_1$, $\omega_2$) vanish. Unlike the rapid conversion scenario, in the slow scenario, the $\omega_i = 0$ points do not necessarily coincide with the critical points of the curves, $\partial M / \partial \epsilon_c = 0$. For our model, all the changes in eigenmodes are from positive to negative values of $\omega_i^2$. In consequence, the stability is never recovered, unlike other works, where a twin star branch appears \citep{Alvarez:2017hmt,Alvarez:2019tfo}. In all cases, the slow stable configurations stay below $\epsilon_c \simeq 6 \times 10^3$~Mev/fm$^3$, corresponding to $n_{B}/n_0 \simeq 24$. This extreme value, impossible to reach in traditional NSs, may allow exploring extreme regions in the QCD phase diagram in the high density-low temperature regime.

\begin{figure}
    \centering
    \includegraphics[width=\linewidth,angle=0]{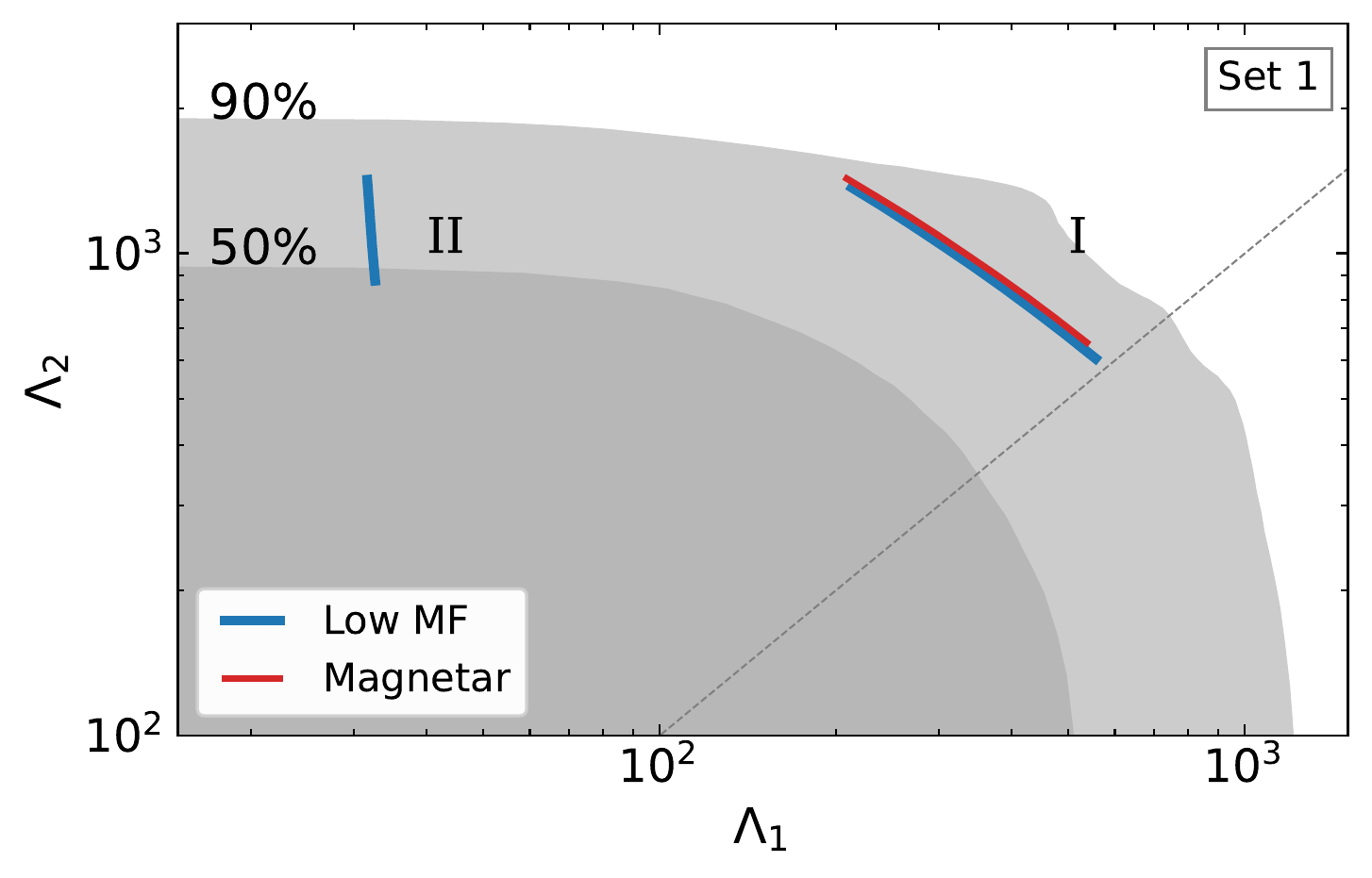}
    \caption{Dimensionless tidal deformabilities, $\Lambda_1$-$\Lambda_2$, for binary NS mergers considering the constraint from GW170817 event \citep{Abbott:2018exr}. Our model, assuming the same chirp mass and mass ratio as GW170817, shows two different possible scenarios. The sets of Table~\ref{table:sets} satisfy this constraint when the binary system is composed of two purely hadronic NS (label I). The results from Set~$1$ also include the possibility of complying with the constraint by a slow HS-hadronic NS binary merger scenario (label II).}
    \label{fig:tidales}
\end{figure}

Also, we study the dimensionless tidal deformability, $\Lambda$, taking into account the astrophysical constraints coming from the GW170817 event. We calculate the $\Lambda$-$M$ relationship, as for a $1.4~M_{\odot}$ NS, the tidal deformability was found to be in the range $70 < \Lambda_{1.4} < 580$ at the $90\%$ credibility level \citep{Abbott:2018exr}. Our model satisfies this constraint regardless of FCM sets; this occurs due to the $1.4~M_{\odot}$ configuration belongs to the purely hadronic branch. Further, in Fig.~\ref{fig:tidales} we show the $\Lambda_1$-$\Lambda_2$ plane, constructed considering the two binary components of this merger event. Our results are within the constrained region through two different scenarios. For all the sets, our model satisfies the constraint in the scenario where both components are purely hadronic (label I). In particular, in a second scenario for Set~$1$, it is also possible to fulfil the constraint if the least massive component is a purely hadronic NS and the most massive is a HS of the extended branch (label II). Therefore, the slow conversion regime allows not only the existence of new kinds of objects but also to accomplish the current astrophysical constraints through this new channel.

\section{Stellar Non-Radial Oscillations}
\label{oscillations}

The general formalism used to study non-radial oscillations due to polar perturbations was carried out in \citet{LD1983} and \citet{DL1985}. The perturbed metric can be described by the line element:
\begin{eqnarray}
\label{pert_ds}
 {\rm d}s^2 &=& - {\rm e}^{\nu} (1+r^\ell H_{0 \ell m} Y_{\ell m} e^{i\omega t}) {\rm d}t^2 \nonumber \\
 && - 2 i\omega r^{\ell+1} H_{1 \ell m} Y_{\ell m} e^{i\omega t} {\rm d}t {\rm d}r
       \\
&& + {\rm e}^{\lambda} (1 - r^\ell H_{0 \ell m} Y_{\ell m} {\rm e}^{i \omega t}) {\rm d}r^2 \nonumber \\
         &&+ r^2 (1 - r^\ell K_{\ell m} Y_{\ell m} {\rm e}^{i\omega t}) {\rm d}\Omega^2 \, . \nonumber
\end{eqnarray}
Moreover, the Lagrangian perturbation to the fluid is described with the following expressions for each of its components:
\begin{eqnarray}
\label{pert_fluid}
\xi^r &=& {\rm e}^{-\lambda /2} r^{\ell -1} W(r)  Y_{\ell m}{\rm e}^{i\omega t}, \nonumber \\
\xi^\theta &=& -r^{\ell-2}V(r) \partial_\theta Y_{\ell m}{\rm e}^{i\omega t},\\
\xi^\phi &=& -\frac{r^{\ell -2} V(r)}{\sin^2\theta}  \partial_\phi Y_{\ell m}{\rm e}^{i\omega t}, \nonumber
\end{eqnarray}
where $Y_{\ell m}$ denotes the spherical harmonics, $\omega$ the frequency of the perturbation, $\nu$ and $\lambda$ the functions that describe the background metric, and $\Omega$ the solid angle.

Particularly, when HSs with discontinuous EoS are considered, it has been shown that it is  useful to define a variable $X(r)$ (see \citet{minuitti2003} for details), using the following algebraic relationship:
\begin{equation}
    X(r) = \omega ^2 (\epsilon + P) {\rm e}^{-\nu /2}V(r) - \frac{{\rm e}^{(\nu - \lambda)/2}}{r}P^\prime W(r) + \frac{\epsilon + P}{2} {\rm e}^{\nu /2} H_0(r), \nonumber
\end{equation}
where the prime denotes a radial derivative.

By adopting this formalism, the perturbations (inside the star) are fully characterised by a system of fourth-order differential equations for the unknowns $X(r)$, $W(r)$, $K(r)$, and $K_1(r)$ that, dropping the $\ell m$ dependence of these 4 unknown functions, are given by:
\begin{eqnarray}
\label{pert_eqs}
H_1^{\,'} &=& -r^{-1} [l+1+2M e^{\lambda}/r + 4 \pi r^2 e^{\lambda} (P-\epsilon)] H_1 \nonumber \\
&& + r^{-1} e^{\lambda} [H_0 + K - 16 \pi (\epsilon+P) V], \nonumber \\
 K^{\,'} &=& H_0/r + \tfrac{1}{2}l(l+1)r^{-1}H_1 - [(l+1)/r-\nu'/2]K  \nonumber \\
&& - 8 \pi (\epsilon+P) e^{\lambda/2} r^{-1} W, \\
 W^{\,'} &=& -(l+1)r^{-1}W + r e^{\lambda/2} [(\gamma P)^{-1} e^{-\nu/2} X \nonumber \\
&& - l(l+1)r^{-2} V + \tfrac{1}{2} H_0 + K], \nonumber \\
 X^{\,'} &=& -l r^{-1} X + (\epsilon+P)e^{\nu/2} \{ \tfrac{1}{2} (r^{-1}-\nu'/2) H_0 \nonumber \\
&& + \tfrac{1}{2} [r \omega^2 e^{-\nu}  + \tfrac{1}{2} l(l+1)/r] H_1 + \tfrac{1}{2} (3 \nu'/2 - r^{-1})K  \nonumber \\
&&- l(l+1) (\nu'/2) r^{-2} V - r^{-1} [4 \pi (\epsilon+P) e^{\lambda/2}   \nonumber \\
&& + \omega^2 e^{\lambda/2-\nu}- r^2 (r^{-2}e^{-\lambda/2} \nu'/2)'] W \}, \nonumber
\end{eqnarray}
where $n= (\ell-1)(\ell+2)/2,$ 
and $\gamma $ is the adiabatic index,
\begin{equation}
\label{gamma_pert}
\gamma=\left.\frac{\epsilon+P}{P}\frac{\Delta P}{\Delta \epsilon}\right|_{\rm s},
\end{equation}
where the variations are performed at a fixed entropy per baryon, $s$.

Functions $V(r)$ and $H_0(r)$, can be deduced from algebraic expressions. We will focus on the quadrupolar perturbations, because they are expected to dominate the emission of gravitational waves.

Outside the star, the perturbation equations reduce to the simple Zerilli second-order differential equation \citep{zerilli1970,fackerell1971,chandrasekhardetweiler1975}. As usual, the numerical values of the complex quasi-normal modes are obtained imposing to the Zerilli function a purely outgoing wave behaviour at infinity (for details related to the numerical method, see Ref. \citet{Tonetto:2020dgm}, and references therein).

The modes associated to the pulsation of the stellar fluid can be classified according to the main restoring force. The $p$-modes (pressure modes) have the pressure gradient inside the star as the restoring force; $g$-modes (gravity modes) are generated by buoyancy in a gravity ﬁeld, such as the one caused by a density discontinuity; these two families are separated in frequency by the $f$-mode (fundamental mode), associated with global oscillations of the fluid. In the case of cold-catalysed compact stars, the $g$-modes are all degenerated to zero frequency unless a sharp transition occurs in its core; these $g$-modes are often called discontinuity $g$-modes. Furthermore, it has been proven that for these modes to appear, a slow conversion between phases must occur \citep{Tonetto:2020dgm}. Therefore, a detection of a discontinuity $g$-mode could be used to determine not only the existence of quark matter in the inner core of a compact object but also to define the nature of the phase transition. This fact could have deep physical implications, giving clues on the unknown value of the hadron-quark surface tension, and also shedding some light on the poorly understood hadron-quark conversion process \citep[see, for example][and references therein]{bombaci2016EPJA,lugones2016EPJA}.

\subsection{Results for Stellar Non-Radial Oscillations}

\begin{figure*}
    \centering
    \includegraphics[width=0.4\linewidth,angle=0]{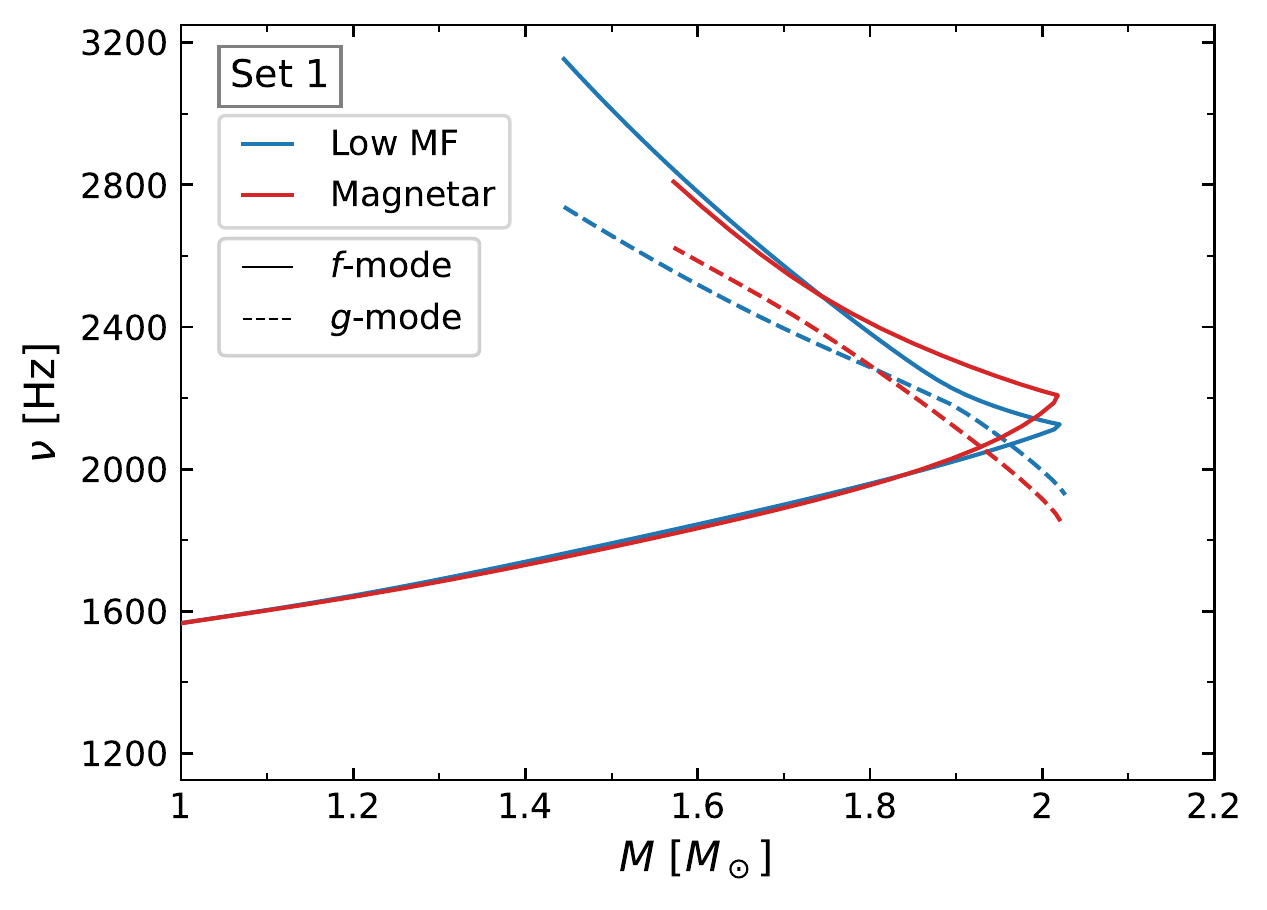}
    \includegraphics[width=0.4\linewidth,angle=0]{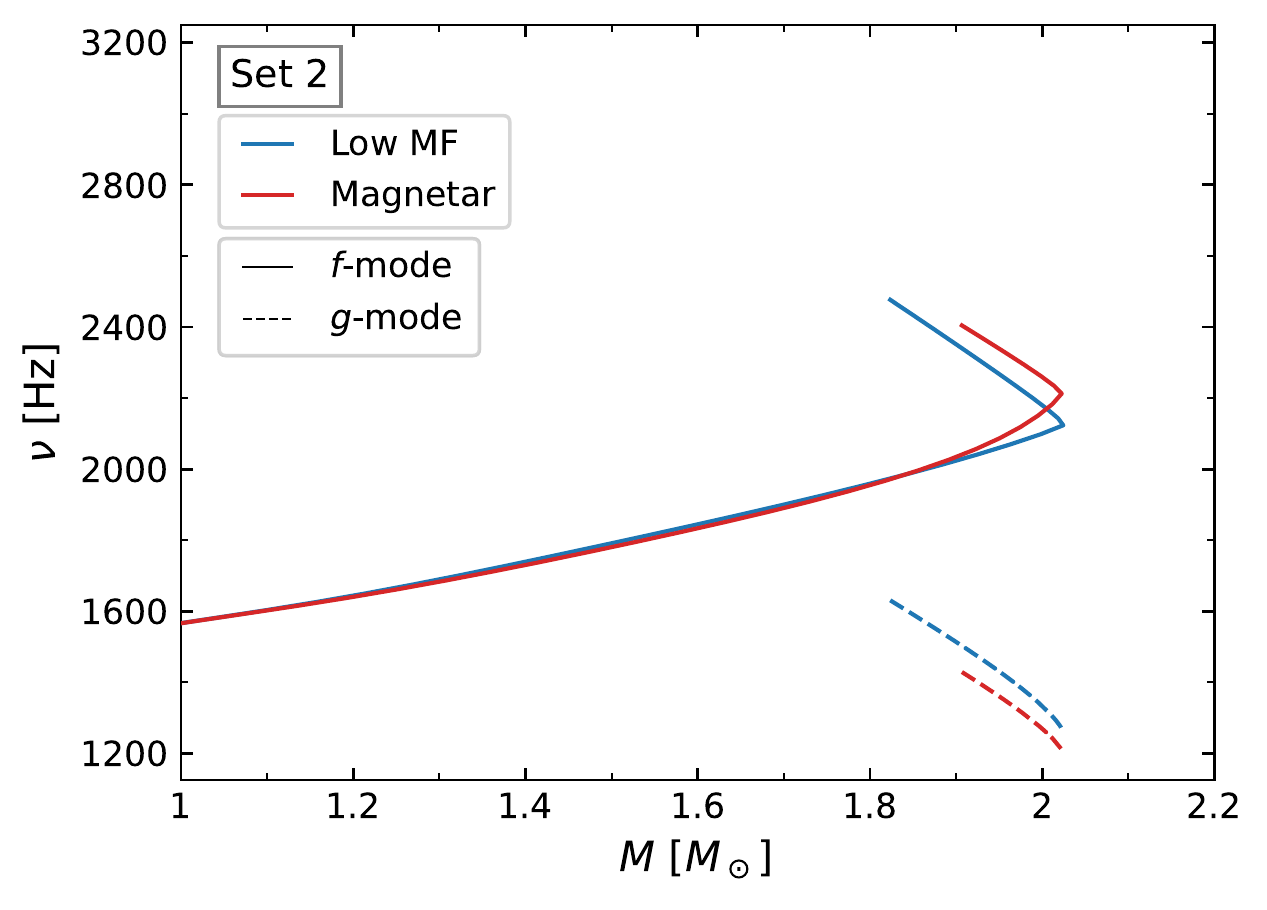}
    \includegraphics[width=0.4\linewidth,angle=0]{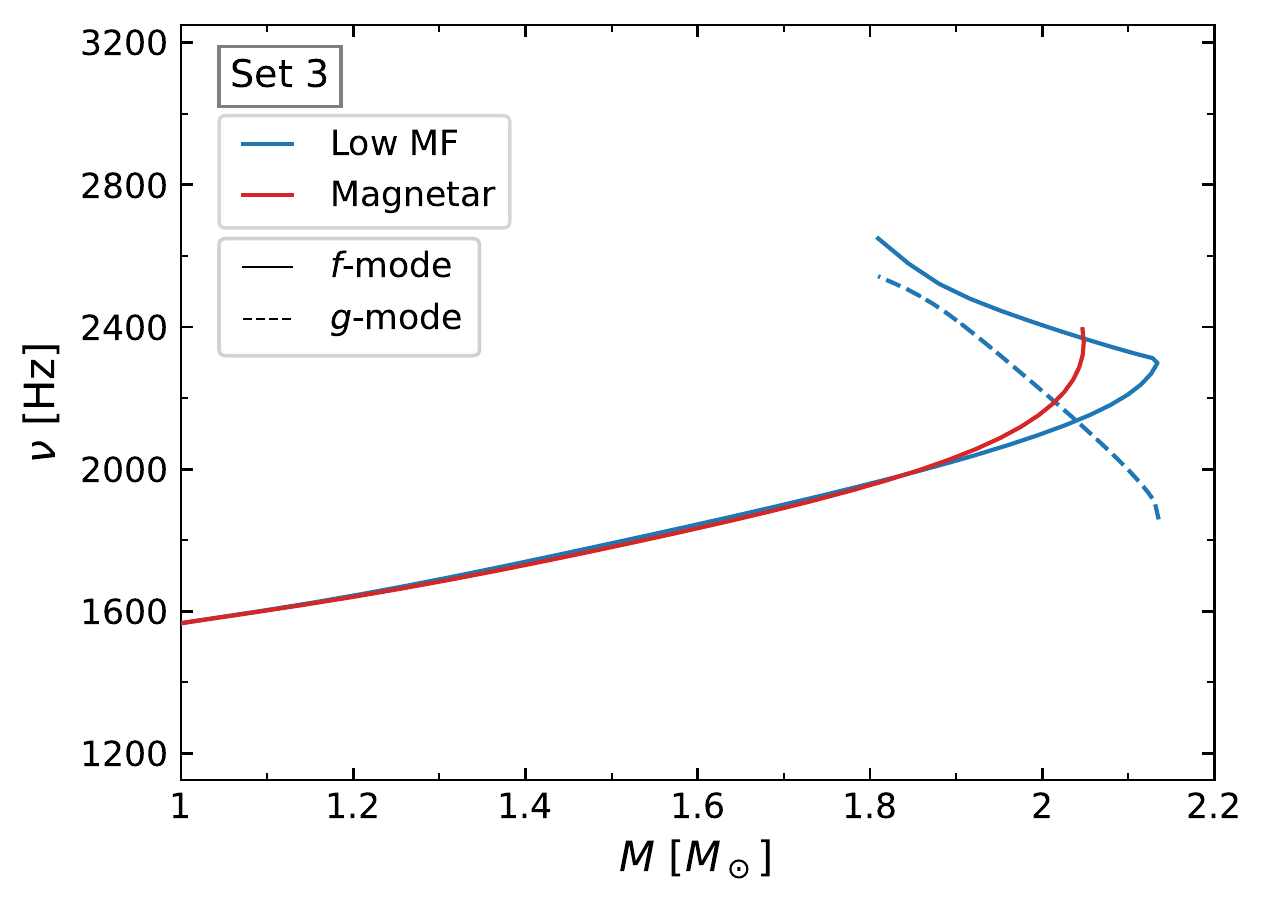}
    \caption{Frequency, $\nu=\textrm{Re}[\omega]/2\pi$, of the $f$ and $g$-modes, as a function of the mass, for the sets of Table~\ref{table:sets}. In the rapid conversion scenario, the $f$-mode curve is valid from lower frequency values only up to the maximum mass, and the $g$- mode does not exist; in the slow conversion scenario, all the continuous line of $f$-modes correspond to stable configurations and the $g$-modes are present. While the traditional values of the $g$-mode frequencies are significantly smaller than those of the $f$-modes, in the slow extended stability branch the frequency values of the $g$-mode are close to or even equal to those of the $f$-mode. Only Set~$2$ shows no overlapping of $f$ and $g$-modes frequencies, being the $g$-mode -not significantly- lower. For the magnetar case of Set~$3$, as there is no existence of extended hybrid branch , the $g$-mode does not appear.}
    \label{fig:freq}
\end{figure*}

\begin{figure*}
    \centering
    \includegraphics[width=0.4\linewidth,angle=0]{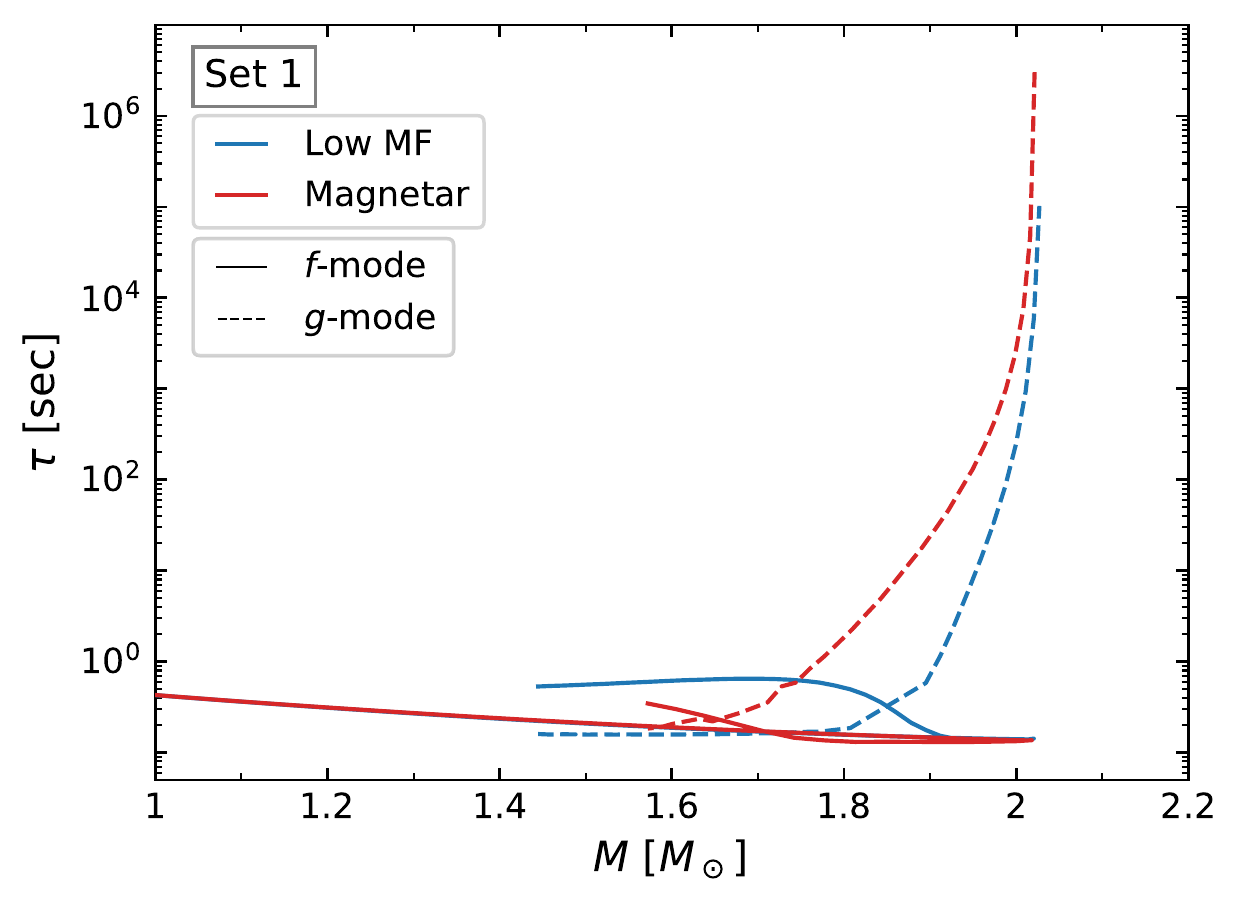}
    \includegraphics[width=0.4\linewidth,angle=0]{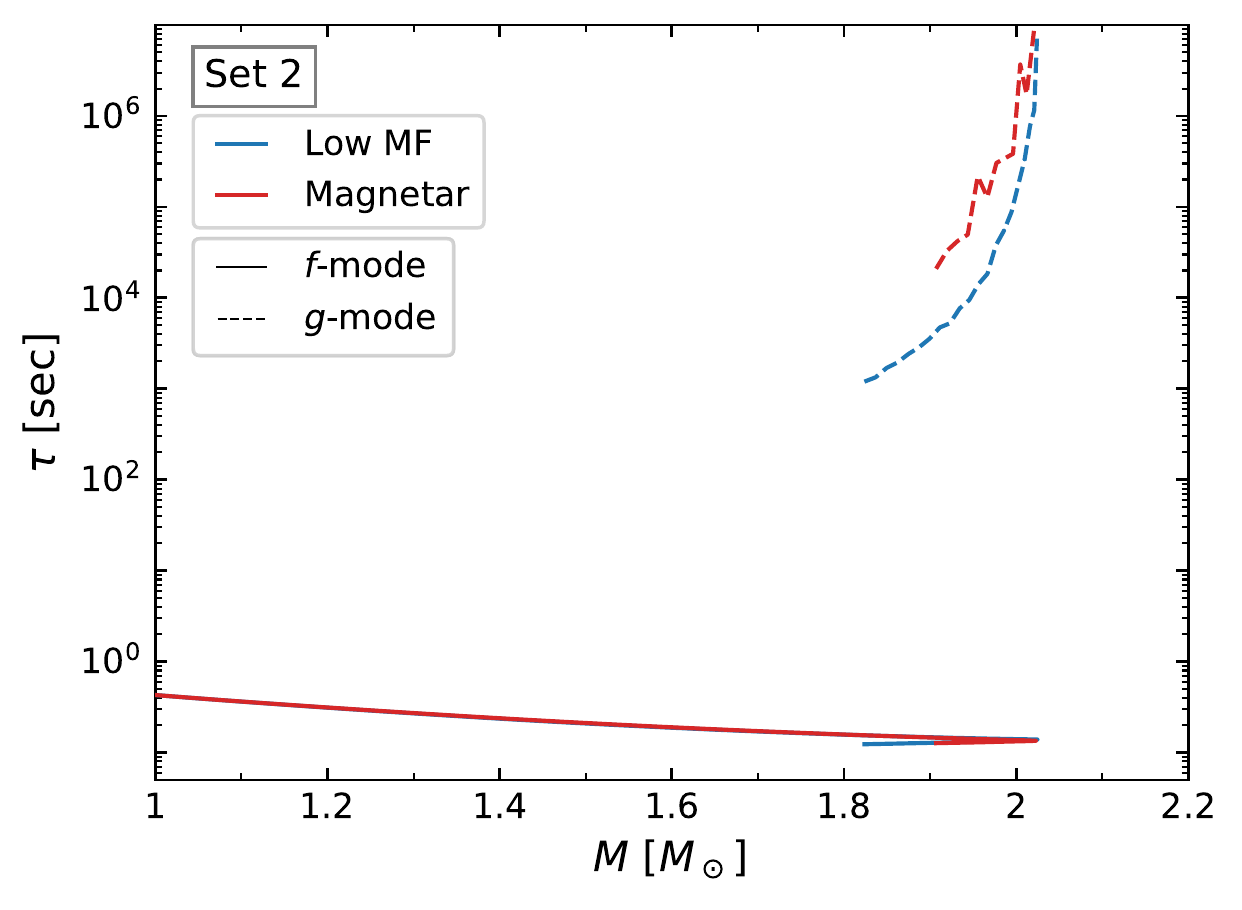}
    \includegraphics[width=0.4\linewidth,angle=0]{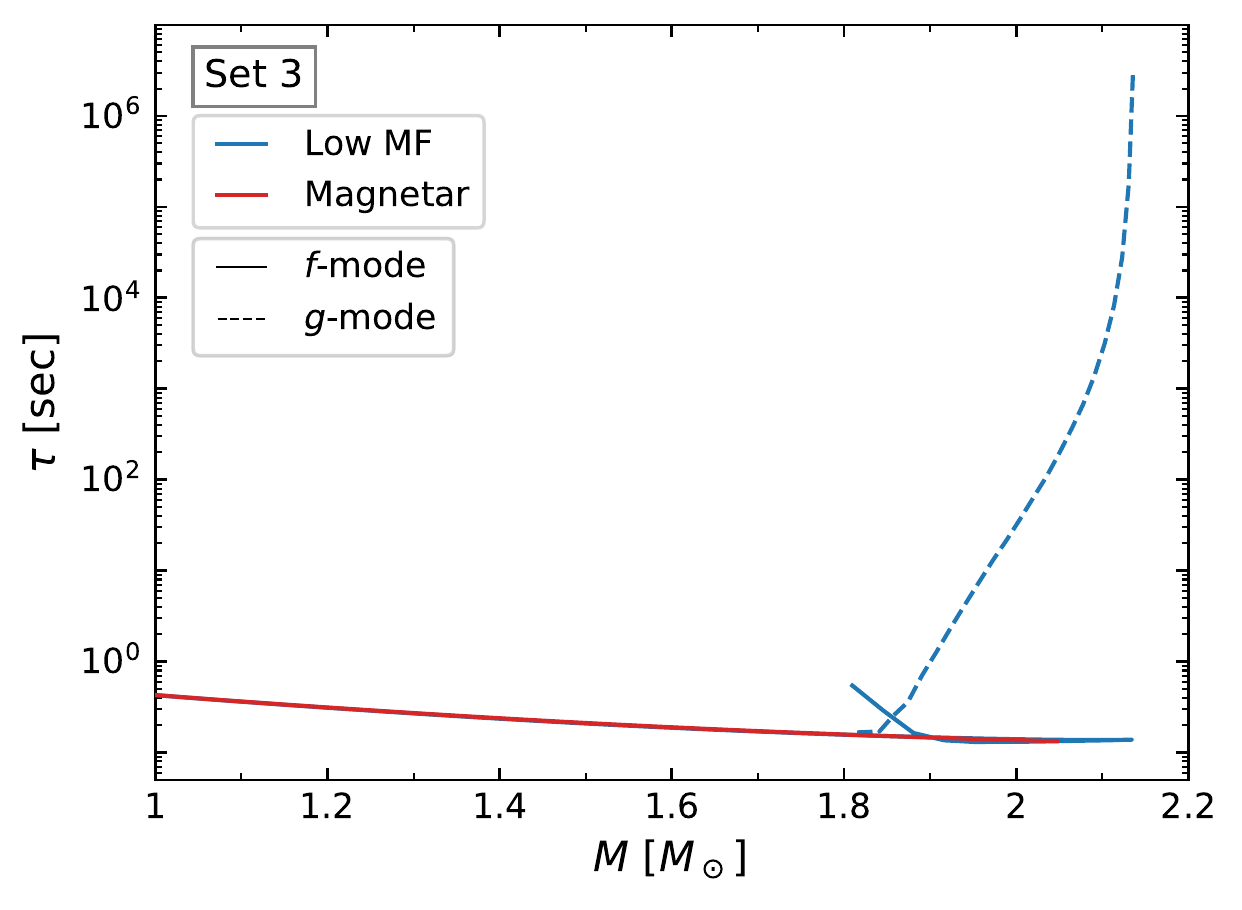}
    \caption{Damping time, $\tau=1/\textrm{Im}[\omega]$, of the $f$ and $g$-modes, as a function of the mass for the sets of Table~\ref{table:sets}. Although the $f$ and $g$-modes damping times differ from each other several order of magnitude, the long extended stability branches partially cancel this feature: $g$-mode damping times for the low masses slow-HSs overlaps with those for $f$-modes. Only the Set~$2$ shows no overlapping between the $f$ and $g$-modes damping times, being the $g$-mode damping time higher by several orders of magnitude.}
    \label{fig:tau}
\end{figure*}

In Figs.~\ref{fig:freq} and \ref{fig:tau}, we present the results of $f$ and $g$-modes for the sets of Table~\ref{table:sets} in the low MF and magnetar scenarios. In Fig.~\ref{fig:freq}, we show the frequencies, $\nu=\textrm{Re}[\omega]/2\pi$, for these modes as a function of the mass. In the slow conversion case, both the extended stability branch -after the maximum mass towards higher frequency values- and the $g$-mode branch appears. In the rapid conversion scenario, only the $f$-mode exists up to the maximum mass star. In agreement with previous results, within the rapid conversion hypothesis (where the appearance of quark matter in the inner core destabilises the stellar configurations), the $f$-modes span in the frequency range $1.5$-$2.2$~kHz. If the phase conversion is slow, the frequency of the fundamental mode increases as the mass decreases up to the terminal mass for objects in the extended stable branch. In some situations, like the low MF case of Set~$1$, the frequencies can be higher than $3$~kHz.

Unlike previous results for $g$-mode frequencies, where the values obtained are notoriously lower than the $f$-mode frequencies \citep{sotani2001PhRvD,minuitti2003,Ranea:2018omo,Rodriguez:2021hsw}, for the extended stability branch $g$-mode and $f$-mode frequencies do not differ substantially from each other; moreover, only for Set~$2$ they do not overlap\footnote{It is worth mentioning that every time we talk about overlapping, it does not mean that the same object has the same frequency or damping time for both modes, but only that the curves cross each other considering different objects in each curve. Consequently, this feature does not affect the avoided crossing effect, but it could make the observational identification or differentiation of the modes difficult.}. The $g$-mode frequencies for all sets are larger than $1$~kHz; in particular, for Set~$1$, $g$-modes present frequencies up to $\sim 2.8$~kHz at the terminal mass. These results are consistent with those obtained in the work of \citet{Tonetto:2020dgm}. The low frequency values for the $g$-modes of Set~$2$ are a consequence of its narrow $\Delta \epsilon$ jump at the phase transition \citep[for more details see][]{Ranea:2018omo,Rodriguez:2021hsw}. It is interesting to note the \emph{avoided crossing} phenomenon of modes in Sets~$1$ and~$3$, in which the presence of the $g$-mode \emph{pushes} the $f$-mode of compact objects in the extended branches to higher frequency values. On the other hand, for Set~$3$, as the magnetar case does not present an extended hybrid branch, the $g$-mode does not appear. Although there are noticeable differences between the frequency modes of low MF and the magnetar scenarios for all sets, such differences are no significant. This would imply difficulties to differentiate both scenarios observationally.

In Fig.~\ref{fig:tau}, we show the corresponding damping times, \mbox{$\tau=1/\textrm{Im}[\omega]$}, as a function of the mass. The considerations regarding the extended stability branch and the $g$-modes also apply to these results. The range of values for the damping times of $f$-modes is $0.1$-$1$~s. In the hadronic branch, the damping time always decreases with the mass. This is not the case for the slow stable HSs, where the behaviour is more complicated and depends on the MF. For magnetars, $\tau$ decreases after the phase transition and then, sometimes increases to a maximum value at the terminal mass (see Set~$1$). For low MF, the situation can be the same (see Set~$3$) or, as in Set~$1$, can show an increase after the transition and a posterior decrease up to the terminal mass. The case of Set~$2$ is different since damping times decrease in the whole extended branch of compact objects. Despite some numerical noise, for $g$-modes the damping time decreases from a maximum value of $\sim 10^6$~s ($\sim 10^7$~s for Set~$2$) in the first object with a quark-matter core to a minimum value at the terminal mass, which is highly dependent on the parameters of the quark EoS.

In this figure it is important to remark that the values of the damping time are shown in a logarithmic scale, unlike the frequency results. In this sense, previous works report differences between the $f$ and $g$-modes of several orders of magnitude \citep[see, for example,][]{sotani2001PhRvD,Tonetto:2020dgm}. In particular, the results from \citet{Tonetto:2020dgm} include the study of extended stability branches for HSs, and the authors do also find these order of magnitude differences. However, for Set~$1$ and $3$, our results show large differences in the damping times only for the high mass part of the extended stability branch; as the mass decreases, the differences are shortened, and the curves overlap in the low mass part of this branch. Moreover, it can be seen that, in a model-dependent range of masses, the damping time for the $g$-mode becomes even smaller than the one associated to the $f$-mode; as a consequence, for some objects in the extended branch, the $g$-mode might have its emission efficiency increased. From the calculations using Set~$1$, we can see that this situation is more evident for the low MF case. For Set~$2$, as it occurs for the frequency values, the damping times present large differences between modes for the whole mass range. For this reason, as both, frequency and damping time of $f$ and $g$-modes, might be similar, is not clear that they could be distinguishable (in the whole compact mass range) from an observational point of view as argued in \citet{Tonetto:2020dgm}. On the other hand, for $f$-mode the differences between low MF and magnetar are not noticeable in the traditional hadronic branch because of the logarithmic scale, and they differ only in the extended branch; for $g$-mode, the differences are more noticeable.

It should be pointed out that, so far, supernova simulations suggest that the $f$-mode is the most efficient GW emitter. However, if one considers the possibility of a hadron-quark phase transition taking place in these compact objects, the $g$-mode may play an important role as well. It would be necessary to take into account also the importance of temperature in determining if the phase conversion is slow or rapid. As discussed in \citep{Tonetto:2020dgm}, in cold stars, slow conversions are more likely to take place, while rapid conversions may occur in hot objects (at temperatures $T \sim \ 20 \mathrm{MeV}$ or even lower) \citep{bombaci2016EPJA}. Thus, in simulations of hot objects considering a first-order phase transition, the nucleation timescale may play a role in how the GW energy is channelled through the modes.

When discussing about the detectability of modes, one should take into account various factors, such as the minimum energy that should be emitted by a given mode, how the energy is channelled through different modes, and other dissipation mechanisms in the astrophysical sources (viscosity, neutrino emission etc.). As a first step in analysing that, here we adopt the same philosophy presented in \cite{Tonetto:2020dgm} to estimate the minimum energy that must be released through a mode in order to be detected by a given GW observatory, according to the formula \citep{anderssonferrari2011,kokkotasapostolatos2001}
\begin{equation}
\begin{aligned}
    \frac{E_\mathrm{GW}}{M_\odot c^2} = & \;  3.47 \times 10^{36} \left(\frac{S}{N} \right)^2 \frac{1+4Q^2}{4Q^2}  \times  \left(\frac{D}{10 \mathrm{kpc}} \right)^2 \times \\
 &  \times \left(\frac{\nu}{1 \mathrm{kHz}} \right)^2 \left( \frac{S_n}{\mathrm{1 Hz^{-1}}} \right) \, .
\end{aligned}    
    \label{eq:energyGW}
\end{equation}
Here, $E_\mathrm{GW}$ represents the energy emitted in the form of GWs, $S/N$ is the signal-to-noise ratio, $Q=\pi \nu \tau $ is the quality factor, $D$ the distance to the source, $\nu$ the frequency, $\tau$ the damping time and $S_n$ the GW detector's noise power spectral density. In \cite{Tonetto:2020dgm} the authors have found that $E_\mathrm{GW}$ is lower for $g$-modes and here we confirm these results. Indeed we have found that the minimum energy emitted by the modes should be of the same order of magnitude for $f$ and $g$-modes. Also, the larger differences presented in $\tau$ for Set~$2$ imply a larger difference in $E_\mathrm{GW}$ between $f$ and $g$-modes. In Table~\ref{table:Egw}, we present the values of the energy for the first (lower central density) and last (higher central density) stable stellar configurations in the hybrid extended branch. There, we assume a galactic event at a distance $D \sim \ 10 \ \mathrm{kpc}$, the detector is the future Einstein Telescope with $S_n^{1/2} \sim 10^{-24} \, \mathrm{Hz}^{-1/2}$ and the event would have $S/N=8$. Of course, if $D$ is larger, the minimum energy required also increases. If we take, as example, an event at the Virgo cluster ($D \sim 15 \ \mathrm{Mpc}$), $E_\mathrm{GW}$ increases by six orders of magnitude when comparing to $D \sim \ 10 \ \mathrm{kpc}$. Note that, as previously discussed, the shorter damping time of the here obtained $g$-modes might favour them as better GW emitters; however, at the same time, the minimum energy that should be emitted to detect them increases. To have a final conclusion on whether a detection is feasible or not, one should analyse the GW emission alongside other dissipation mechanisms.

\begin{table}
\centering
\begin{tabular}{|ccc|cc|}
\hline
\multirow{2}{*}{}                            & \multirow{2}{*}{}         & \multirow{2}{*}{Mode} & \multicolumn{2}{c|}{$E_{GW}$ {[}$10^{45}$ MeV{]}} \\
                                             &                           &     & First & Last  \\ \hline
\multicolumn{1}{|c|}{\multirow{4}{*}{Set 1}} & \multirow{2}{*}{Magnetar} & $f$ & 1.94  & 3.13  \\
\multicolumn{1}{|c|}{}                       &                           & $g$ & 1.36  & 2.73  \\ \cline{2-3}
\multicolumn{1}{|c|}{}                       & \multirow{2}{*}{Low MF}   & $f$ & 1.80  & 3.95  \\
\multicolumn{1}{|c|}{}                       &                           & $g$ & 1.48  & 2.98  \\ \cline{2-5} 
\multicolumn{1}{|c|}{\multirow{4}{*}{Set 2}} & \multirow{2}{*}{Magnetar} & $f$ & 1.94  & 2.29  \\
\multicolumn{1}{|c|}{}                       &                           & $g$ & 0.58  & 0.81  \\ \cline{2-3}
\multicolumn{1}{|c|}{}                       & \multirow{2}{*}{Low MF}   & $f$ & 1.79  & 2.43  \\
\multicolumn{1}{|c|}{}                       &                           & $g$ & 0.63  & 1.06  \\ \cline{2-5} 
\multicolumn{1}{|c|}{\multirow{4}{*}{Set 3}} & \multirow{2}{*}{Magnetar} & $f$ & -     & -     \\
\multicolumn{1}{|c|}{}                       &                           & $g$ & -     & -     \\ \cline{2-3}
\multicolumn{1}{|c|}{}                       & \multirow{2}{*}{Low MF}   & $f$ & 2.10  & 2.79  \\
\multicolumn{1}{|c|}{}                       &                           & $g$ & 1.37  & 2.57  \\ \hline
\end{tabular}
\caption{Minimum energy $E_\mathrm{GW}$, Eq.~\eqref{eq:energyGW}, that should be emitted by the $f$ and $g$-modes in order to detect them for the hybrid stars in the extended branch. We consider the Einstein Telescope detector, with a power spectral density of $S_n^{1/2} \sim 10^{-24} \, \mathrm{Hz}^{-1/2}$, and that the event takes place at a distance $D \sim \ 10 \ \mathrm{kpc}$, producing a signal-to-noise ratio $S/N=8$. The columns "First" and "Last" refer to the first and last stellar stable configuration in the extended hybrid branch, respectively, with respect to the maximum mass configuration.}
\label{table:Egw}
\end{table}
 
\section{Summary, Discussion and Conclusions}
\label{conclus}

In this work, we have studied oscillating magnetised hybrid stars, assuming an abrupt phase transition in their cores. We have constructed the magnetised hybrid EoS and we have obtained significant astrophysics quantities related to the structure, stability, composition, and oscillating properties of these objects. We have approached two relevant NS scenarios: the classic low MF NS and the magnetar. To construct the hybrid EoS, we have included a modern magnetised sub-nuclear EoS for the crust. For the hadronic phase, which includes the baryonic octet and $\Delta$-baryons, we adopt the density dependent SW4L parametrization, satisfying both the nuclear constraints and the recent astrophysical determinations of mass and radius that the NICER Collaboration has done. To describe the quark phase composed of $u$, $d$ and $s$ quarks we use the FCM model. Conditions of $\beta$-equilibrium and local charge neutrality were imposed to the hybrid EoS. Different scenarios for the nucleation timescale of the hadron-quark conversion, which has strong implications on the dynamical stability of the stellar configurations, also were considered. In addition, we have studied the effects of the MF on the EoS microphysics, considering the Landau level quantisation of the electrically charged particles and the AMM of neutral particles. We have adopted the chaotic MF hypothesis due to the disordered distribution of MF field lines inside the star, and we have modelled the MF strength through a parametric exponential function.

We have studied the pressure anisotropy due to the local privileged direction of the MF. There are two quantities contributing to the anisotropy: the magnetisation pressure and the pure magnetic term. Our results have shown that the magnetisation pressure is only relevant for extremely high MF. Hence, it is considered only for the quark phase. We have found that the main contribution to the differences on the pressure components comes from the pure magnetic term; this contribution reaches its maximum value in the hadron phase before the phase transition and, for higher densities, the difference between the pressure components progressively decreases. Within the chaotic MF hypothesis, such differences are only local and they do not have global effects on the star symmetry. Therefore, assuming the spherical symmetry is preserved is a good approximation for the studied stellar configurations. 

We have included the AMM only in the energy spectrum of neutral particles, following the work of \citet{Ferrer:2015iot}. In our treatment of the MF, we have worked consistently using the linear expression for the AMM in the self-energies of the hadron matter particles when it corresponds. As \citet{Ferrer:2015iot} did for charged particles, we found that neutral particle AMM does not significantly affect EoS results, as \cite{Ferrer:2019ton} already suggest for an only-neutron EoS; hence, its contribution to the astrophysical quantities we are interested in is negligible, unlike other authors claimed \citep{Mao:2003aso, Casali:2014hah,Franzon:2015ass}. Therefore, we only present astrophysical results for the low MF and magnetar scenarios, excluding the AMM. Although for the macroscopic results the effects of AMM is irrelevant, its contribution substantially modify the fraction of $\Delta^0$ particles in the hadronic phase. This might have a direct impact on the cooling of HSs, opening up dUrca processes involving these particles.

We have selected three particular sets of the FCM parameters to present the astrophysical results. These sets were chosen to be qualitatively representative of the general behaviour of our model and to satisfy the restriction of $2.01~M_\odot$ and the current astrophysical constraints coming from GWs and X-ray observatories. After this choice, we have solved the TOV equations taking into account rapid and slow hadron-quark conversion regimes and their impact on the dynamical stability. The maximum mass configurations and the length of the extended stability branches show differences between the three sets considered. In particular, Set~$1$ presents the longest extended stability branch, allowing the possibility of explaining the high-mass component of GW170817. This feature is also present in the $\Lambda_1$-$\Lambda_2$ plane, where Set~$1$ satisfies the dimensionless tidal deformability constraints including the possibility of an HS-hadronic NS merger event. Thus, the possible existence of an extended stability branch due to the slow conversion regime could allow satisfying the current astrophysical constraints in a novel way. In this sense, besides our particular EoS model, if the hybrid extended branch exists and is long enough, it would be possible to satisfy the GW170817 event with objects from this branch, and the traditional hadronic branch could be allowed to have larger radii. This possibility should be further investigated in the future.

It is important to note at this point, that, in our model, HSs only exist if the slow conversion scenario is taken into account. In the rapid scenario, stable HSs exist only marginally because the appearance of quark matter in their cores induces a critical point in the $M(\epsilon_c)$ relationship. In the slow conversion scenario, extended stability branch is always hybrid. Therefore, the detection of a star belonging to this branch would confirm the existence of an abrupt phase transition in the core of NSs.

The maximum mass peak in the curves of the HSs families does not always increase as MF increases, as \citet{Flores:2020gws} also obtained using a different EoS and contrary to other works
\citep{Rabhi:2009qhp,Mariani:2019mhs,Thapa:2020eos,Rather:2021hmn}. This suggest a model dependent behaviour. In particular, for Set~$3$, the maximum mass decreases as the MF increases for the magnetar scenario, compared with the low MF case, and the difference in the maximum mass peak between both curves is $\Delta M_{\rm max} \sim -0.1M_\odot$. 

We have also implemented an improved magnetised crust EoS presented in \citep{Mutafchieva:2019rol}. We have found that given a hadronic EoS, low mass stars, $M \lesssim 1~M_\odot$, are more compact than in our previous work \citep{Mariani:2019mhs}, where we have used the magnetised crust of \citet{Lai:1991ceo}.

The MF also affects the radius of the HSs and the length of the extended stability branch. For magnetars, the extended branch is always shorter than for low MF stars. Within our model, a particular point in the $M$-$R$ plane appears, $M \simeq 1.9 M_\odot$ and $R \simeq 12$~km, where for all the sets the low MF and magnetar curves cross each other. Magnetars have larger radii than low MF HSs for configurations with lower masses than this point, and smaller radii for stable configurations with higher masses. 

We have also calculated the frequencies and damping times for the non-radial $f$ and $g$ oscillation modes. Our results show that the differences in the frequencies of both modes for low MF HSs and magnetars are noticeable, but not far enough to distinguish them with near future GWs detectors. In the slow conversion scenario, the extended hybrid branch has remarkable implications on the frequencies of both modes, showing an increase on their values for the slow stable configurations. This increase also produces an approaching between the frequency curves of $f$ and $g$-modes, even overlapping in some cases; for our results, only Set~$2$ does not show overlapping, having a shorter $g$-mode frequency branch. However, not always the occurrence of a hybrid extended stability branch produces an abrupt increase on the $g$-mode frequency values; comparing our results with those presented by \citet{Ranea:2018omo,Rodriguez:2021hsw}, this increase is an EoS dependent feature. As it was demonstrated in \citet{Ranea:2018omo,Rodriguez:2021hsw}, the value of the $g$-mode frequency is strongly related to the energy density discontinuity at the phase transition, $\Delta \epsilon$. In our work, Set~$1$ and $2$ have larger values of $\Delta \epsilon$, so that, predictably, the frequencies of the $g$-mode are comparable to those of the $f$-mode.

As the differences between damping times usually range several orders of magnitude, the damping times of low MF stars and magnetars might be more feasible to differentiate observationally than frequencies. For $f$-mode damping times, the extended stability branch also separates from the traditional branch and might be distinguishable in a future observation. Comparing our results with those presented by \citet{Tonetto:2020dgm}, the authors obtain that damping times of $f$ and $g$-modes are always separated several orders of magnitude, as it happens for Set~$2$ in our work. However, for Sets $1$ and $3$, we obtain overlapping between modes for damping times, as it can be seen from Fig.~\ref{fig:tau}. We have obtained that, for these two sets, some $g$-modes of the extended branch have damping times even smaller than those of the $f$-mode. For this reason, the GW-dissipation mechanism of this mode might become as effective as the $f$-mode. Considering the possible observation of a damping time, if the detection occurred for an object close to the maximum mass configuration, the possibility of differentiating between scenarios or modes would increase. As it was pointed out in \citet{Tonetto:2020dgm}, the $g$-mode from slow extended stability branches has, in general, significantly larger frequencies sand smaller damping times than the related to compact objects in the traditional branches, and so that they could become more feasible detected for current and planned GWs observatories, assuming a extended stability branch exist. We have also estimated the minimum energy that should be emitted by a mode in order to detect it with the future Einstein Telescope detector. As the damping times of $g$-modes in the extended hybrid branch becomes significantly smaller than the usual ones, this makes this minimum energy to increase. Therefore, to draw a firm conclusion on whether a $g$-mode detection is feasible or not, one should analyse how the GW energy is channelled through the modes and take into account other dissipation mechanisms such as viscosity and neutrino emission.

To conclude, all the results presented in our work point in the same direction: in the multi-messenger era with GWs, observable quantities suitable to constrain and learn about the microphysics aspects of compact objects exist. Moreover, some key aspects of the hadron-quark phase transition properties, such as the surface tension or the nucleation timescale, might be indirectly determined through some observable predictions we obtained. Although our results only show subtle differences between the cases and scenarios presented, advances in measurements and restrictions on NSs obtained in recent years began to elucidate some of these questions. We expect that the predictions about magnetised HSs, differences between low MF NSs and magnetars, and the slow and rapid conversions could be observationally addressed and tested in the future.

\section*{Acknowledgements}

The authors thank the anonymous referee for the constructive comments and criticisms that have contributed to improve the manuscript substantially. The authors want to acknowledge N. Chamel and his research group, Z. Stoyanov, and Y. Mutafchieva for contributing to this work with the magnetised crust EoS calculations. M.M. and M.C.R are fellows of CONICET. M.M., I.F.R.-S. and M.G.O. thank CONICET and UNLP (Argentina) for financial support, under grants PIP-0714 and G157, G007. I.F.R.-S. is also partially supported by PICT 2019-0366 from ANPCyT (Argentina). I.F.R.-S., M.G.O. are supported by the National Science Foundation (USA) under Grant PHY-2012152. L.T. thanks the Italian Istituto Nazionale di Fisica Nucleare (INFN) under grant TEONGRAV. A.P.M. thanks Agencia Estatal de Investigación through the grant PID2019-107778GB-100 from Junta de Castilla y Leon, Spanish Consolider MultiDark FPA2017-90566-REDC and PHAROS COST Actions MP1304 and CA16214. 

%%%%%%%%%%%%%%%%%%%%%%%%%%%%%%%%%%%%%%%%%%%%%%%%%%
\section*{Data Availability}

The computed data presented and discussed in this paper will be shared upon reasonable request.



\bibliographystyle{mnras}
\bibliography{biblio} % if your bibtex file is called example.bib

%\begin{thebibliography}{99}
%\end{thebibliography}

%%%%%%%%%%%%%%%%%%%%%%%%%%%%%%%%%%%%%%%%%%%%%%%%%%

%%%%%%%%%%%%%%%%% APPENDICES %%%%%%%%%%%%%%%%%%%%%

%%%%%%%%%%%%%%%%%%%%%%%%%%%%%%%%%%%%%%%%%%%%%%%%%%

% Don't change these lines
\bsp	% typesetting comment
\label{lastpage}
\end{document}